\documentclass[sn-vancouver]{sn-jnl}

\usepackage{graphicx}%
\usepackage{fancyhdr}
\usepackage{multirow}%
\usepackage{amsmath,amssymb,amsfonts}%
\usepackage{amsthm}%
\usepackage{mathrsfs}%
\usepackage[title]{appendix}%
\usepackage{xcolor}

\usepackage{xcolor}%
\usepackage{textcomp}%
\usepackage{manyfoot}%
\usepackage{booktabs}%
\usepackage{algorithm}%
\usepackage{algorithmicx}%
\usepackage{algpseudocode}%
\usepackage{mathtools}  

\usepackage{physics}
\usepackage{caption}
\usepackage{subcaption}
\usepackage{braket}
\usepackage{tikz}
\usetikzlibrary{quantikz2}
\usepackage{multirow}
\usepackage{hyperref}
\usepackage[parfill]{parskip}
\usepackage{listings}%
\usepackage{multicol}
\usepackage{float}
\usepackage{placeins}

\begin{document}

\title[Fourier Series Guided Design of Quantum Convolutional Neural Networks for Enhanced Time Series Forecasting]{Fourier Series Guided Design of Quantum Convolutional Neural Networks for Enhanced Time Series Forecasting}


\author*[1]{\fnm{} \sur{Sandra Leticia Ju\'arez-Osorio}}\email{sandra.juarez@cinvestav.mx}

\author[1]{\fnm{} \sur{Mayra Alejandra Rivera-Ruiz}}\email{mayra.rivera@cinvestav.mx}

\author[1]{\fnm{} \sur{Andres Mendez-Vazquez}}\email{andres.mendez@cinvestav.mx}

\author[2]{\fnm{} \sur{Eduardo Rodriguez-Tello}}\email{ertello@cinvestav.mx}

\author[3]{\fnm{} \sur{Jos\'e Mauricio L\'opez-Romero}}\email{jm.lopez@cinvestav.mx}

\affil[1]{\orgdiv{}, \orgname{CINVESTAV Unidad Guadalajara}, \orgaddress{\street{ Av. del Bosque 1145}, \city{Zapopan}, \postcode{45019}, \state{Jalisco}, \country{M\'exico}}}

\affil[2]{\orgdiv{}, \orgname{CINVESTAV Unidad Tamaulipas}, \orgaddress{\street{Km. 5.5 Carretera Victoria - Soto La Marina}, \city{Ciudad Victoria}, \postcode{87130}, \state{Tamaulipas}, \country{M\'exico}}}

\affil[2]{\orgdiv{}, \orgname{CINVESTAV Unidad Quer\'etaro}, \orgaddress{\street{Libramiento Norponiente 2000, Fracc. Real de Juriquilla}, \city{Santiago de Quer\'etaro}, \postcode{76230}, \state{Quer\'etaro}, \country{M\'exico}}}


\abstract{In this work, we apply 1D quantum convolution in the task of time series forecasting. By encoding multiple points into the quantum circuit to predict subsequent data, each point becomes a feature and the problem becomes a multidimensional one. Taking as basis previous theoretical works which demonstrated that Variational Quantum Circuits (VQCs) can be expressed as multidimensional Fourier series, the capabilities of different architectures and ansatz are explored. This analysis considers the concepts of circuit expressivity and the presence of barren plateaus. Analyzing the problem within the framework of the Fourier series enabled the incorporation of data reuploading in the architecture, resulting in enhanced performance. Rather than a strict requirement for the number of free parameters to exceed the degrees of freedom of the Fourier series, our findings suggest that even a limited number of parameters can produce Fourier functions of higher degrees. This highlights the remarkable expressive power of quantum circuits. This observation is also significant in reducing training times. The ansatz with greater expressivity and number of non-zero Fourier coefficients consistently delivers favorable results across different scenarios, with performance metrics improving as the number of qubits increases.\\}


\maketitle

\section{Introduction}\label{sec1}

Today, Quantum Computing (QC) is still in the Noisy intermidiate-Scale Quantum (NISQ) era: devices with a small number of qubits and errors. Variational Quantum Circuits work with few qubits, being suitable to this kind of devices \citep{variational}. These VQCs are capable of working in conjunction with classical layers in order to divide the task between a classical and a quantum computer. Several applications to these hybrid algorithms can be found in the literature (\cite{variational,surveyQML,review}). In this work, we focus  on Quantum Machine Learning, specifically in the utilization of Variational Quantum Circuits (VQCs) to construct a quantum version of the widely used Convolutional Neural Networks (CNN's) \citep{henderson}. \\

In the literature, numerous instances abound where Quantum Machine Learning (QML) algorithms have demonstrated superior or comparable performance to their classical counterparts. This holds true across various scenarios, encompassing toy datasets (\cite{PhysRevA.101.032308,park2022variational,abejas}) as well as real-world applications, such as medical image classification (\cite{sameer2022novel,10.1093/jcde/qwac003,alzhimer}), defects detection in materials \citep{defects}, and time series forecasting (\cite{time_series,mayra}). Notably, Quantum Convolutional Neural Networks (QCNNs) have found application not only in toy datasets like MNIST and Fashion MNIST (\cite{qconvolution,henderson}), but also in practical contexts, such as protein distance prediction \citep{hong2021quantum}, and the identification of COVID-19-infected patients \citep{10.1093/jcde/qwac003}. \\

While some of these works adopt an empirical approach to achieve superior results compared to classical counterparts, our focus in this study is to provide insights based on existing theory, specifically  from (\cite{fourier,fourier2,expressibility,barren}). Details of each of those works are provided in the following sections. The theoretical foundation described in those works is helpful to establish a general framework to design an efficient architecture that yields both good metrics and manageable training times.\\

In \cite{fourier} and \cite{fourier2}, a demonstration of how VQCs can be viewed as Fourier Series (FS) is depicted, establishing a relationship between the number of layers, data reuploading, and the expressivity of the circuit. \cite{expressibility} introduces a measure of ansatz expressivity, while \cite{barren} discusses the challenge of training circuits with flat cost landscapes.\\

 The main contributions of this work can be summarized as follows. To the best of our knowledge we present the first empirical study that validates the impact of the parameters introduced in \cite{barren}, and \cite{expressibility} when training with real data, utilizing them to design an effective architecture. We work with time series forecasting, which is a problem that naturally can take advantage from visualizing the VQCs as FS. Focusing specifically in time series, our work acknowledges that training times can be quite prohibitive for large data sets. The importance of designing the architecture in the context of the FS will be assessed in terms of the relation between the improved expressivity of the circuits and the flat cost landscapes. Our findings demostrate  a starting point for designing architectures with greater expressivity even with a smaller number of quibits involved on the architecture. This is of great importance for designing efficient architectures in quantum computing given the actual constraints in the number of qubits.\\
 
 \textcolor{black}{
The analysis of the 1D QCNN within the theoretical framework presented in the work of \cite{fourier} led to the design of an improved model architecture. Additionally, we found that the quantum circuits were able to achieve comparable performance metrics using fewer trainable parameters than those predicted by the theory presented in \cite{fourier2}, highlighting the expressive power of quantum models.\\
}

 The rest of the work is organized in the following way. First in section 2, a general background is presented. In this part, we briefly introduce  quantum computing, VQC's and the theory of VQC's as Fourier series from \cite{fourier}  and \cite{fourier2} is presented. Next, the  1D  quantum convolution \citep{mayra} is explained and how it can be seen from the context of multidimensional Fourier series. Also we describe the expressivity as in \cite{expressibility} and the measure of barren plateaus as in  \cite{barren}. In section 3 we explain the architectures that will be tested in this work, the datasets the description of the simulations. In the results, section 4, we discuss the relevance of reuploading data, the number of necessary trainable parameters and the different results obtained by utilizing various architectures, ansatz and number of qubits. Finally, a conclusion is presented in the last section. 
 
\section{Background and related work}
\label{background}
\subsection{Quantum Computing}

The emergence of quantum computers, proposed by Feynman in 1982 for simulating quantum systems, has evolved significantly in the past four decades, despite  accuracy limitations in quantum processors (\cite{feynman2018simulating,40years}). Quantum algorithms have supremacy in certain problems: For example  Shor's algorithm for factoring and Grover's search algorithm \citep{nielsen}. In contrast to classical computers, which follow deterministic laws of physics, microscopic quantum systems when isolated from their environment, exhibit non-classical behaviors such as uncertainty, collapse, and entanglement (\cite{nielsen,40years}).\\

Now, lets briefly introduce some concepts of quantum computing. Analogously to the bit in classical computing, the quantum bit or qubit is the basic unit of information processing used in quantum computing.  Unlike classical bits, which can only exist in states of 0 or 1, qubits follow the principles of superposition and entanglement. In superposition, a qubit can simultaneously occupy both states 0 and 1 until measured, providing a potential computational advantage over classical bits \citep{nielsen}. In quantum mechanics,  qubits and their interactions are mathematically represented as vectors of the Hilbert space. The state space of $n$ qubits is described as a tensor product space, enabling the representation of complex quantum states and operations \citep{nielsen}.\\

Quantum gates are unitary operators utilized to perform transformations on qubits. These gates, similarly to classical logic gates, perform operations on qubits, preserving the quantum information encoded within them. With the sequential application of quantum gates, quantum circuits manipulate qubit states to perform specific computational tasks. Various quantum gates, including Pauli matrices and the Hadamard gate are utilized in circuits to perform operations on qubits, enabling the creation of entangled states and the implementation of quantum algorithms. These gates, together with control operations such as the CNOT and CZ gates, form the basis of quantum circuits \citep{nielsen}. Those last kind of gates are utilized to generate entangled states. Multiple advantages on having entagled states can be mentioned, but particularly in the context of VQCs and Quantum Machine Learning, having a circuit with a strong entangling structure is capable to better cover the Hilbert space and to capture correlations in data (\cite{expressibility}).  \\

\textcolor{black}{
Several quantum-inspired models have proposed using tensor network architectures to efficiently capture complex feature correlations in high-dimensional spaces. For example, Matrix Product States (MPS) have been used in supervised learning tasks to compress input representations while retaining expressivity \cite{stoudenmire2016supervised}. Other approaches, such as those by Huggins et al., explore the implementation of tensor network structures like MPS and Tree Tensor Networks (TTNs) on quantum hardware \citep{huggins2019towards}. More recently, the Residual Tensor Train (ResTT) model extends classical tensor train architectures with skip connections to capture feature interactions of multiple orders while improving training stability through mean-field analysis \citep{chen2022residual}. While these models differ in design, they share with our work the conceptual motivation of using structured, physically inspired representations to capture rich correlations while maintaining computational efficiency. However, our approach departs significantly in methodology: instead of relying on classical tensor decompositions or fixed unitary constructions, we leverage variational quantum circuits (VQCs) with parameterized quantum gates, trained in a hybrid quantum-classical loop. This allows us to flexibly model frequency components guided by Fourier theory, while directly targeting NISQ-compatible quantum implementations.
}

\subsection{Variational Quantum Circuits}
\begin{figure*}[]
	\centering
	\includegraphics[scale=0.45]{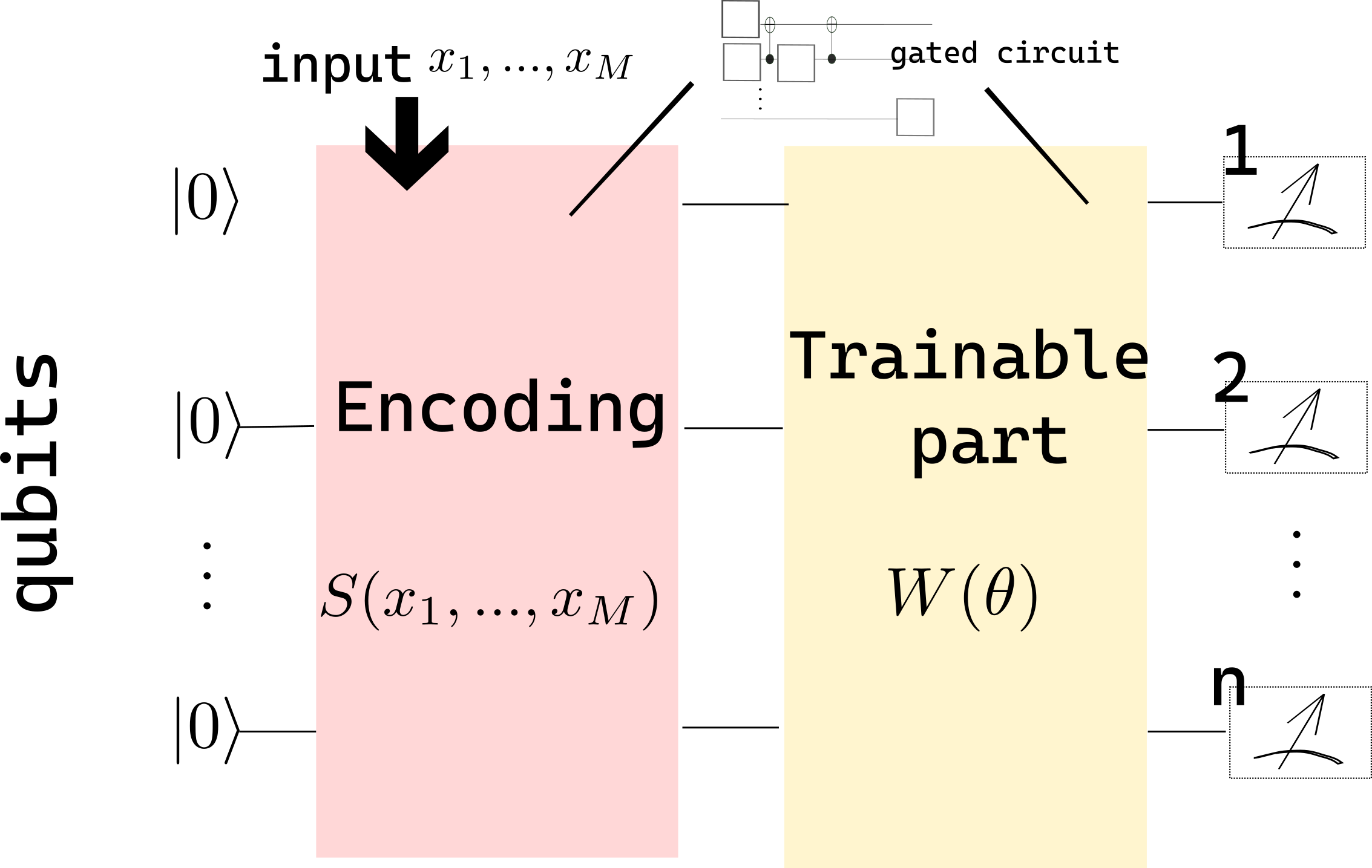}
	\caption{Variational Quantum Circuit (VQC's). The classical input $\mathbf{x}$ is encoded into a quantum state $\ket{\Psi_{\mathrm{in}}(\mathbf{x})}$ with the unitary transformation $U_{\mathrm{in}}(\mathbf{\boldsymbol{x}})$ applied to the initial quantum state $\ket{0}^{\otimes n}$. In the second step, a unitary transformation $U(\mathbf{\boldsymbol{\theta}})$ with trainable parameters is applied. Finally, a measurement is made on the qubits.}
	\label{vqc}
\end{figure*}

Variational Quantum Circuits are trainable quantum circuits that are widely used as quantum neural networks for different tasks. The general structure of a VQC is presented in Figure \ref{vqc}. VQCs are quantum algorithms that capture correlations in data using entangling properties \citep{PhysRevA.101.032308}. In today's noisy intermediate-scale quantum computers (NISQ), which suffer from noise and qubit limitations, the VQC is the leading strategy due to their shallow depth \citep{variational}.\\

At Figure \ref{vqc}, the first step is to encode the classical input $x$ into a vector in the Hilbert space. This is accomplished by applying a unitary transformation $S_{\mathrm{in}}(x)$ to the initial state, which is generally chosen as $\ket{0}^{\otimes n}$ \citep{variational}. Several strategies are utilized in the encoding stage and two common approaches are:

\begin{itemize}
\item \textbf{Amplitude encoding:} The concatenated input of $M$ inputs and $N$ features $x(x_1^{1},...,x_N^{1},...,x_1^{1},...,x_N^{1})^{T}$  is associated with probability amplitudes of a quantum state \citep{libro_supervised}. 
\item \textbf{Rotation encoding: } This method embeds a classical point $x_i$ into a single qubit as $\ket{\phi(x)}$ in the following way:  $U_{\phi}(x): x \in \Re^{N}  \longleftrightarrow \ket{\phi(x)}\\
= \bigotimes_{i=1}^{N}\left( \cos\left(\frac{x_i}{2}\right)\ket{0} + \sin\left(\frac{x_i}{2}\right)\ket{1} \right)$ 
\end{itemize}

  After encoding the classical input, the state vector is passed through a set of quantum operations $W(\boldsymbol{\theta})$ depending on an optimizable parameter $\theta$ \citep{variational}. The form of the ansatz depends on the specific task; although some ansatz architectures are generic. The parameters $\boldsymbol{\theta}$ can be encoded in a unitary $W(\theta)$ applied to the input states, and the way this parametrization is defined can have a significant impact on training since it defines the shape of the cost function \citep{centric}. The quality of an ansatz can be determined by its expressivity and its entangling capability. It is said that an ansatz is expressible if the circuit can explore the entire space of quantum states \citep{variational}. After passing through the encoding and trainable unitary transformations $U$  the final state would be given as:
\begin{equation}
\ket{\Psi}=U\ket{0}^{\otimes n},
\end{equation}
Now, considering that the initialized state is $\ket{0}^{\otimes n}=(1,0,\cdot \cdot \cdot 0)^{T}$, then the ith element of this state would be given as:

\begin{equation}
\Ket{\Psi_i}=U_{ij}\delta_{j1}=U_{i1}
\label{delta}
\end{equation}

\subsection{VQCs as Fourier series}
A strategy presented in \cite{fourier}  and \cite{fourier2} is to have $L$ layers, reuploading the encoded data each time. A layer consists of the encoding gates $S(x)$, with $x$ in this context a classical input, and a trainable layer $W(\theta)$. The encoding $S(x)$  is composed of gates of the form $e^{ixH}$, with $H$ a Hamiltonian. This can always be decomposed as $H=V^{\dag}\Sigma V$, with $\Sigma$ a diagonal operator with the eigenvalues $\lambda _i$ of $H$. The $ij$  element of the transformation matrix after $L$ layers can be written as:

\begin{equation}
U_{ij}=  \sum_{j_1,\dots,j_L=1}^{N}W^{(L+1)}_{i j_{L}}e^{(ix\lambda_{j_L})}
 \times W^{(L)}_{j_{L}j_{L-1}} \dots W^{(2)}_{j_2j_1}
\times e^{(ix\lambda_{j_1})}W^{(1)}_{j_{1}j},
\label{melement}
\end{equation} 
with $N=2^{n}$ for $n$ qubits. Finally, combining Eq. (\ref{delta}) and Eq. (\ref{melement}) the following result is obtained,

\begin{equation}
\ket{\Psi}_i=\sum_{j_1,\dots,j_L=1}^{N}e^{(i(\lambda_{j_1}+\dots+\lambda_{j_L})x)} \times W^{(L+1)}_{ij_L}\dots W^{(2)}_{j_2j_1}W^{(1)}_{j_11}.
\label{psii}
\end{equation}

In a VQC a physical quantity represented by the observable $M$, is measured as the quantum state passes through the circuit. This measurement yields the expected value, expressed as:

\begin{equation}
\langle M \rangle= \bra{\Psi}M\ket{\Psi},
\label{Eq. measurement}
\end{equation}

and typically the Pauli $Z$ operator is used. The circuit is run multiple times ($S$ repetitions), and the expected value is obtained by averaging measurements. This outcome is associated with labels by assigning probabilities, and a cost function is computed for parameter optimization (\cite{variational,penny,shift}).\\

Finally, combining the Eq. (\ref{psii}) and Eq. (\ref{Eq. measurement})  and using the multi-index $\Vec{j}=\{j_1,...,j_L\} \in [N]^{L}$ and the sum of eigenvalues for $\Vec{j}$ as $\Vec{\Lambda_j}=\lambda_{j_1}+\cdot \cdot \cdot +\lambda_{j_L}$, the result after the measurement is:

\begin{equation}
f(x)=\sum_{\Vec{k},\Vec{j}\in [N]^{L}} c_{\Vec{k},\Vec{j}}e^{{i(\Lambda_{\Vec{k}}-\Lambda_{\Vec{j}})}},
\label{fourier1d}
\end{equation}

with,

\begin{equation}
c_{\Vec{k},\Vec{j}}=\sum_{i,i'} W^{*(1)}_{1j_{1}}W^{*(2)}_{j_1j_2} \cdots 
 \cdots W^{*(L)}_{j_{L}j_{L-1}}W^{*(L+1)}_{j_{L}i}
 \times M_{ii'} W^{(L+1)}_{i' j_{L}}W^{(L)}_{j_{L}j_{L-1}} \cdots   \cdots W^{(2)}_{j_2j_1}W^{(1)}_{j_{1}j}.
\label{coefficients}
\end{equation}

Now, if the terms with the same frequency $\omega=\Lambda_k-\Lambda_j$ in the sum in (\ref{fourier1d}) are grouped, a Fourier series is obtained \cite{fourier,fourier2}:

\begin{equation}
f(x)=\sum_{\omega}c_{\omega}e^{i\omega x}.
\end{equation}
In this last expression, the coefficients are the summation of all the $ c_{\Vec{k},\Vec{j}}$  that contribute to the same frequency and as it can be seen in Eq. (\ref{coefficients}). They only depend on the trainable and measurement part of the circuit.  On the other hand, the frequency spectrum is related to the eigenvalues of the gates of the encoding part. As it was proven in \cite{fourier}, the accessible frequency spectrum of a circuit can be linearly incremented by reuploading the encoding $L$ times: Repeating a Pauli encoding $L$ times supports a spectrum of size $L$. Also, they show how different structures for the ansatz in the trainable part can produce different subsets of Fourier coefficients and set some coefficients to zero.  The frequency spectrum and the variety of the coefficients are directly related to the expressivity of the circuit. For example, as shown in \cite{fourier}, a single Pauli rotation encoding is only capable of learning a sine function since the model's frequency spectrum only consists of a single non-zero frequency. By increasing the spectrum of frequencies they show that the circuit is capable of learning more complex target functions. The flexibility in the coefficients is hard to investigate: Each trainable block and the measurement part contributes to every Fourier coefficients \citep{fourier}, hence a small change in  $W(\theta)$ is capable to produce a complete different set of coefficients.  \\


\subsection{1D Quantum convolution as multidimensional Fourier series }
\label{1dqc}

\begin{figure*}[b]
	\centering
	\includegraphics[scale=0.4]{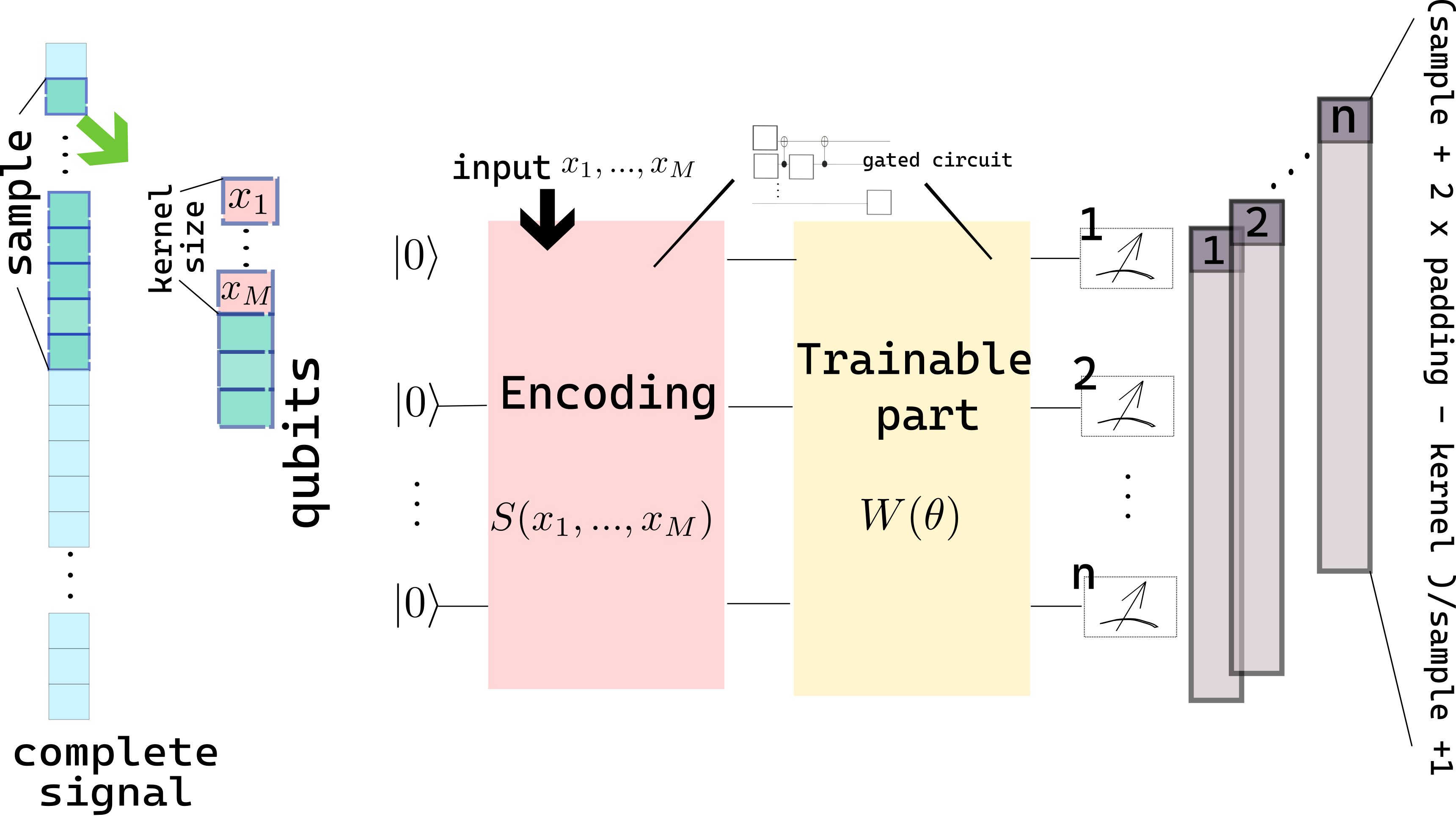}
	\caption{1D quantum convolution as introduced in \cite{mayra}, which utilizes a VQC. This circuit has $n$ qubits and encodes $M$ features. The output is a matrix with classical values.}
	\label{1Dquanvolution}
	\end{figure*}

The 1D quantum convolution proposed in \cite{mayra}  takes the approach presented in \cite{henderson}, but with the difference that is adapted to one dimension, and the quantum layer is trainable.  Instead of employing element-wise matrix multiplication as in the classical convolution, the 1D quantum convolution processes subsections of one-dimensional signals using a VQC, with the structure presented in  Figure \ref{vqc}, to generate a feature map. The schema of the  1D Quantum Convolution proposed in \cite{mayra} is depicted in Figure \ref{1Dquanvolution}. It takes as input a sliding window of size $k$ of  a sequence of size $s$. This input is  encoded in the quantum circuit. After this,  a the trainable part of the VQC is applied. Subsequently, information is decoded through measurements on all qubits, yielding real-number outputs (Eq. \ref{Eq. measurement}). This output is a matrix with dimensions $(n,o)$, with $o=(s+2+p-k)/r+1$, with $p$ the padding and $r$ the stride. Details on the specific architecture, number of qubits and classical layers utilized in this work are discussed in Section 3. \\

As the outputs of the 1D quantum convolution consist of real component vectors, this quantum layer can be  incorporated into diverse architectures, whether quantum or classical. In a study by \cite{mayra}, a hybrid approach is explored by integrating a 1D quantum convolution with a classical 1D convolution and linear layers. The model's performance is benchmarked against a purely classical counterpart, resulting in superior metrics for the hybrid configuration. In contrast, our focus in this work is to assess and compare the capabilities of different quantum architectures in achieving optimal outcomes. In this work, 1D quantum convolution is applied to the problem of time series forecasting: Multiple points are encoded into the quantum circuit to predict the next one. A deeper explanation about the complete model is presented in Section 3. \\

Since the problem treated in this work involves the encoding of multiple points, it is necessary to introduce the multidimensional version of the Fourier analysis presented in the previous section. In \cite{fourier2} an extension of \cite{fourier} is presented where they analyze use of a multidimensional by proposing several architectures generating a multidimensional truncated Fourier series given by the following equation,

\begin{equation}
f(\Vec{x})=\sum_{\omega_1,...,\omega_M=-D}^{D}c_{\Vec{\omega}}e^{{i\Vec{x} \cdot\Vec{\omega}}}.
\end{equation}

Here, the vector element $D=\max(\omega_1,...,\omega_M)$ is the degree of the Fourier series, $\Vec{x}$ is an $M$ dimensional vector, and the complex coefficients $c_{\Vec{\omega}}$ have the property $c_{\omega}=c^{*}_{-\omega}$.\\

The time series that will be utilized in this work are unidimensional, but the problem can be treated as multidimensional since several points are needed to predict the next point. Each time step can be seen as a dimension, representing a feature, so encoding multiple points involves working with a multidimensional feature space in which each point contributes to a different dimension in this feature space. \\

The degrees of freedom $\nu$ of a Fourier series are the number of necessary independent variables to characterize a set of coefficients for a certain degree $D$ and are given by the following expression \citep{fourier2}: 

\begin{equation}
\nu=(2D+1)^{M}.
\label{nu}
\end{equation}

\begin{figure*}[t]
\centering
\includegraphics[scale=0.3]{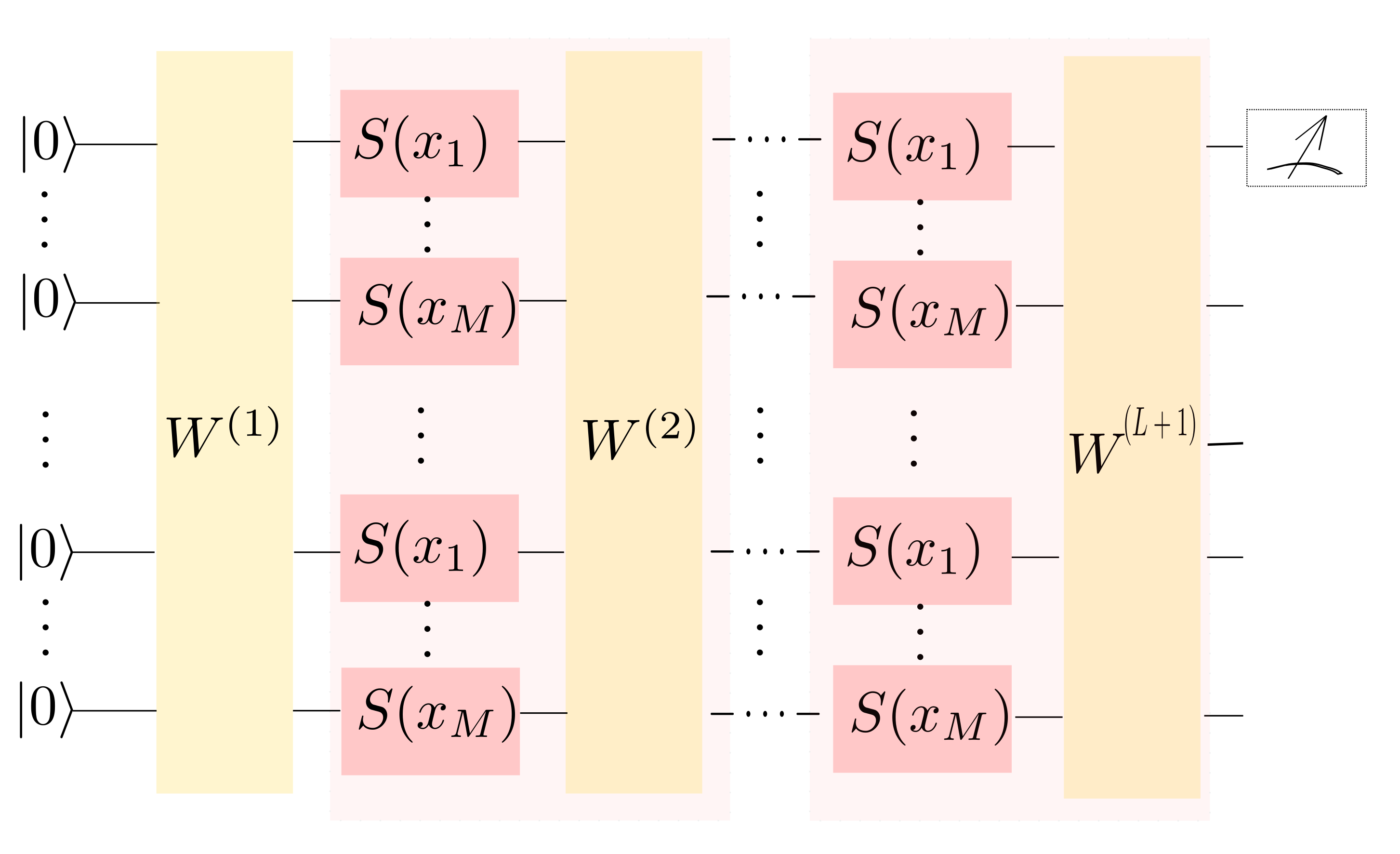}
\caption{Super parallel architecture proposed in \cite{fourier2} which outputs a multidimensional Fourier series. It encodes $M$ features which are reuploaded vertically and horizontally. The outputs are classical values.}
\label{sp}
\end{figure*}

As a minimum condition, the proposed architectures are required to have more or an equal number of trainable parameters than the degree of freedom to have a general enough circuit. In \cite{fourier2} several architectures are proven to output a multidimensional Fourier series, but only the one called \textit{super parallel ansatz} always satisfies the condition of having more free parameters than the degrees of freedom for any degree $D$.   This architecture, illustrated in Figure \ref{sp}, consists of stacking layers in depth and width directions. It requires $n=LM$ qubits and each layer can be represented as:

\begin{equation}
L(\Vec{x},\Vec{\theta})=\left( \bigotimes_{l=1}^{L}\left( \bigotimes_{m=1}^{M}S(x_m)\right)\right)W^{(l)}(\Vec{\theta}),
\end{equation}
with $l$ the lth repetition of blocks vertically. The number of free parameters for this model is $N_p=(L+1)(2^{2ML}-1)$, considering that, as it is proposed in \cite{fourier2}, the unitary gates composing the trainable part have $N^2-1$ trainable parameters. This approach represents difficulties in computing times at the training stage, as it will be discussed in the following section.\\

\subsection{Expressivity}
In \cite{expressibility} the expressivity is described as a circuit's ability to generate states representative of the Hilbert space. They propose to calculate it by comparing the distribution of the states generated by the circuit with the distribution of the Haar random states. \\

Given a dimension $N$, the unitary matrices of $N\times N$ constitute the unitary group $U(N)$. The Haar measure tells how to weight elements of $U(N)$ when performing operations with the unitary group \citep{measeure}. An ensemble of Haar random states or $t$-design means that the first $t$ moments of the Haar measure can be exactly captured: reproducing higher moments better approximates the full unitary group. Now,  a frame potential is defined as:

\begin{equation}
\mathcal{F}^{(t)}=\int_\theta \int_\phi \norm{\bra{\psi_{\theta}}\ket{\psi_\phi}}^{2t}d\Vec{\theta}d\Vec{\phi}.
\end{equation}
For an ensemble of Haar random states, $\mathcal{F}^{(t)}=\frac{t!(N-1)!}{(t+N-1)!}$ with $N=2^{n}$ for $n$ qubits. This value is a lower bound for the frame potentials: $\mathcal{F}^{t}\geq \mathcal{F}^{(t)}_{Haar}$. In \cite{expressibility} it is observed that potentials can be seen as the distribution of state overlaps, $P(F=\norm{\bra{\psi_{\theta}}\ket{\psi_\phi}}^{2})$, with $F$ the fidelity. The analytical form of the probability density function of fidelities is known for the ensemble of Haar random states \citep{measeure}:

\begin{equation}
P_{\mathrm{Haar}}(F)=(N-1)(1-F)^{N-2}.
\end{equation}

 In \cite{expressibility}  the fidelity distribution is estimated by sampling pairs of states to obtain the probability distribution of fidelities with a histogram. With this, the expressivity is calculated with the Kullback-Leibler divergence between the estimated fidelity distribution and that of the Haar-distributed ensemble \citep{expressibility}:

\begin{equation}
E=D_{KL}(P_c(F) | P_{\mathrm{Haar}} (F))
\end{equation}
A lower KL divergence with respect to the Haar distribution corresponds to a more expressible circuit. The upper bound of expressivity is given by $(N-1)\ln(b)$, with $b$ the number of bins in the histogram \citep{expressibility}. \\

In the context of \cite{expressibility}, an expressive ansatz is capable to explore uniformly the space of a unitary group. Also, an expressive ansatz is said to be problem agnostic, it can be utilized to solve a wide range of problems. Then, this indicates that the expressivity of a VQC can also be assessed in the context of the FS, explained Subsection 2.3 and 2.4, analyzing the frequency spectrum and accessible coefficients. As previously mentioned, when more frequencies and coefficients are accessible to a particular circuit, more complicated functions can be approximated by the model, meaning that a wider range of problems can ve solved. In this work both ways of analyzing the expressivity of a circuit are compared. 

\subsection{Barren plateaus}

A circuit is said to exhibit a barren plateau if the gradient vanishes exponentially when adding more qubits, leading the optimizer to be trapped in a local minimum. The average of $\partial_k C$, with $C$ the cost function over the set of trainable parameters  $\boldsymbol{\theta}$ is zero, meaning that the gradient cannot take any direction, or what is known as vanishing gradient. In \cite{barren}, the trainability of an ansatz is assessed with the Chebyshev inequality, which bounds the probability that the partial derivative deviates from its average of zero:
\begin{equation}
P(\abs{\partial_kC}\geq\delta)\leq \frac{\mathrm{Var}[\partial_kC]}{\delta^{2}}.
\end{equation}
From this equation, it can be interpreted that when the variance is small, the gradient vanishes with high probability and those cases are untrainable. In \cite{barren}, it is found that expressive ansatz exhibits barren plateaus. 

\section{Proposed model}
\label{proposed_model}

\begin{figure}[t]
	\centering
	\begin{subfigure}[b]{0.93\textwidth}
		\centering
		\includegraphics[width=0.93\linewidth]{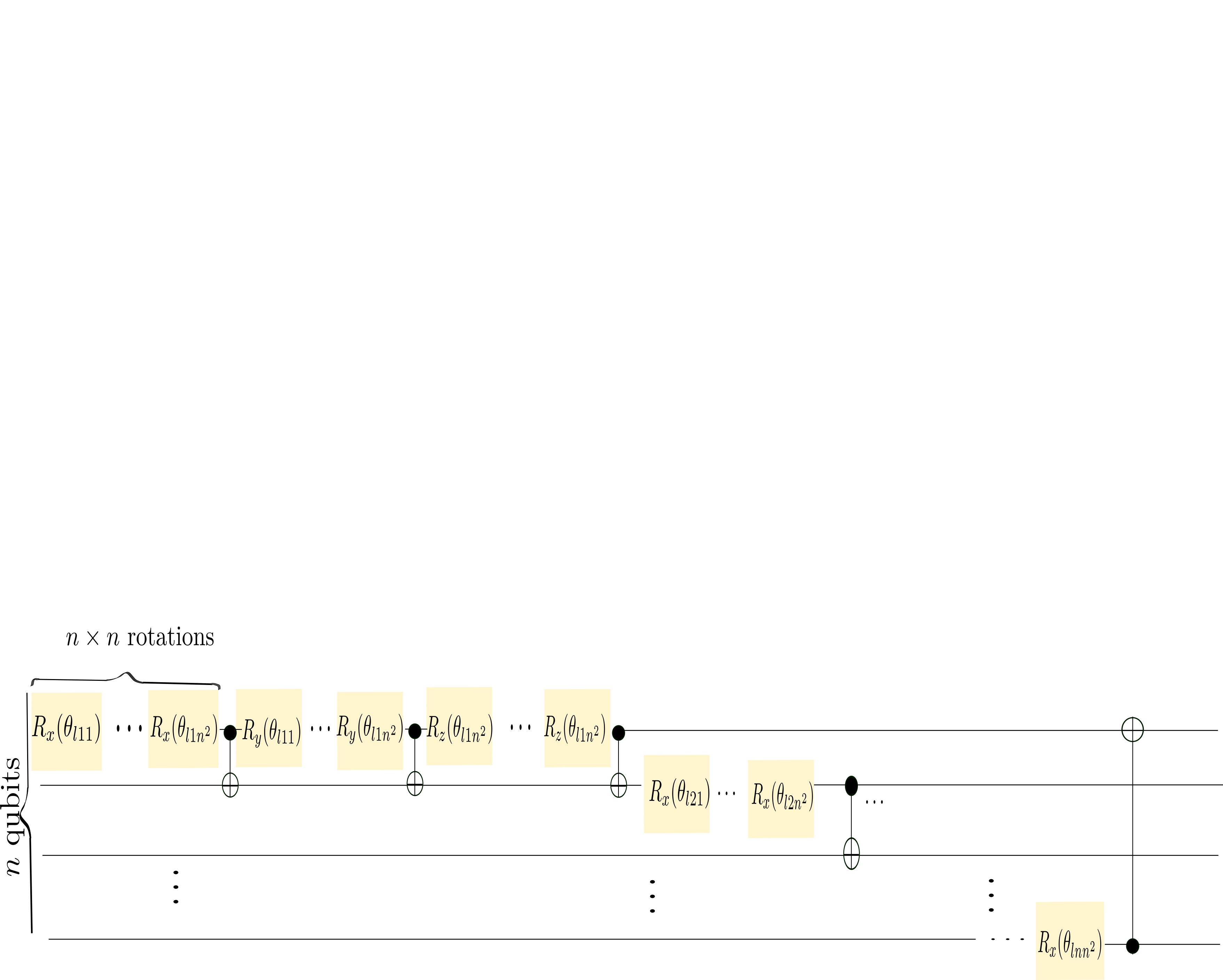}
		\caption{$n\times n=192$ trainable parameters.}
		\label{Anxn}
	\end{subfigure}%
	
	\begin{subfigure}[b]{0.90\textwidth}
		\centering
		\includegraphics[width=0.90\linewidth]{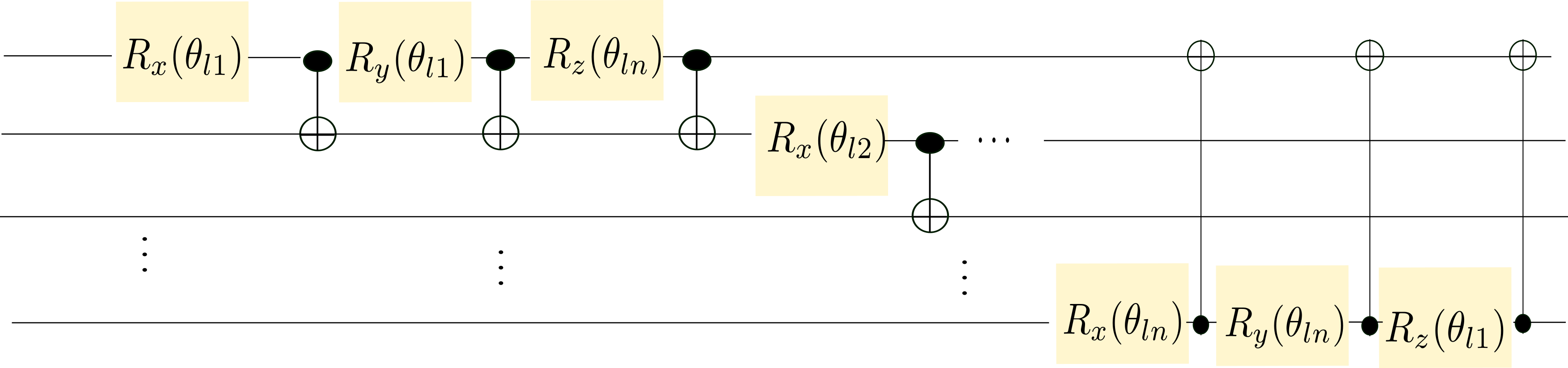}
		\caption{$n\times L=12$ trainable parameters.}
		\label{Acustom}
	\end{subfigure}
	
	\begin{subfigure}[b]{0.32\textwidth}
		\centering
		\includegraphics[width=0.77\linewidth]{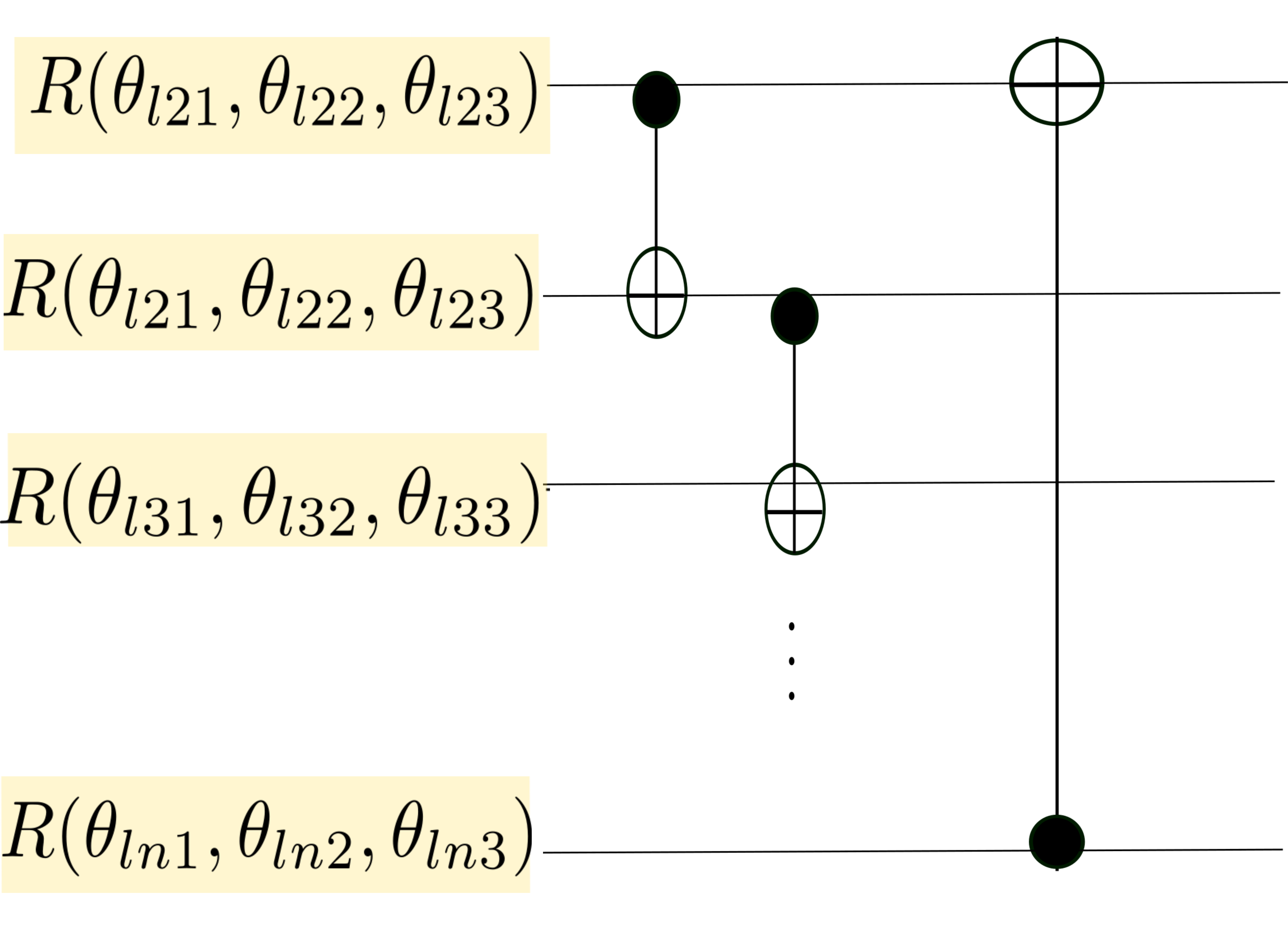}
		\caption{$3\times n \times L=36$ trainable parameters.}
		\label{Abasic}
	\end{subfigure}%
	\hfill
	\begin{subfigure}[b]{0.32\textwidth}
		\centering
		\includegraphics[width=0.75\linewidth]{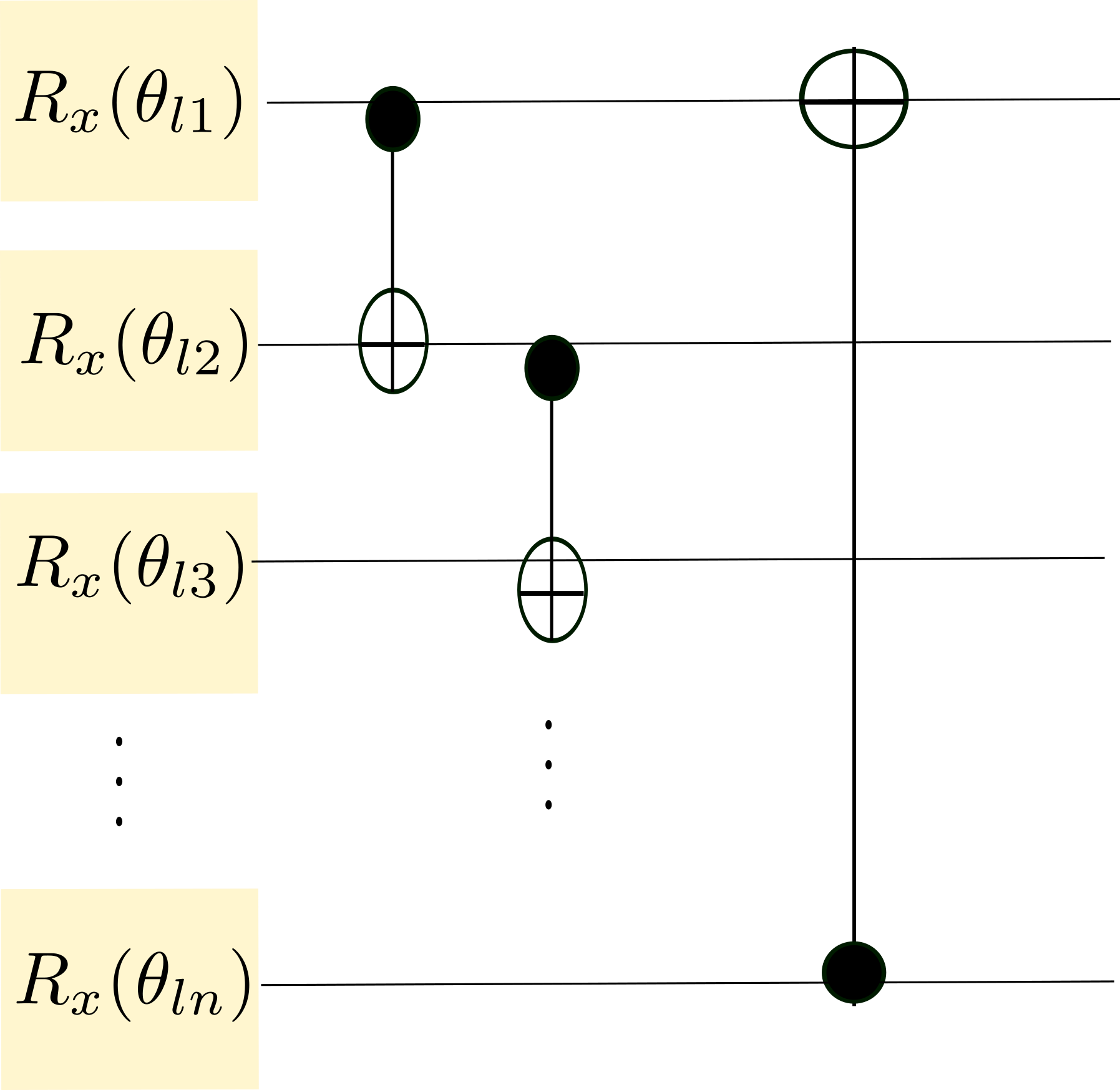}
		\caption{$n\times L=12 $ trainable parameters.}
		\label{Astrongly}
	\end{subfigure}%
	\hfill
	\begin{subfigure}[b]{0.32\textwidth}
		\centering
		\includegraphics[width=0.92\linewidth]{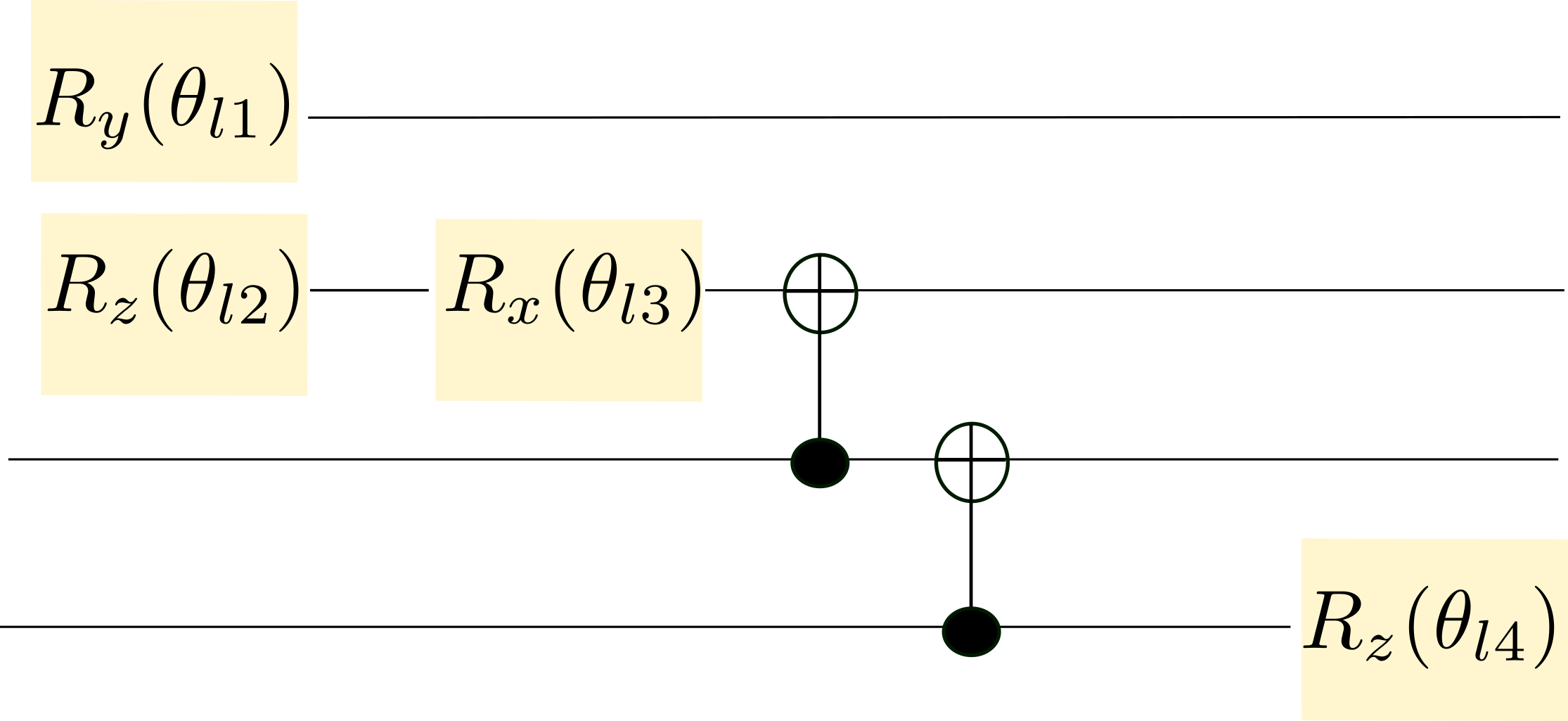}
		\caption{$n \times L=12$ trainable parameters.}
		\label{Arandom}
	\end{subfigure}
	\caption{Ansatz for the trainable part $W(\theta)$. Here $n$ is the number of qubits. This repeats in each layer $l$ of the circuit.}
	\label{architectures}
\end{figure}

In this work the \textit{super parallel ansatz} depicted in Figure \ref{sp} is the general schema to follow, varying the part $W(\Vec{\theta})$ to compare the performance both in metrics and in the subset of coefficients that each choice of $W(\Vec{\theta})$ is capable to achieve. As mentioned above, utilizing gates of $N^2-1$ trainable parameters would imply large training times. For this reason, in this work the choice of $W(\Vec{\theta})$ will be layers with only $n\times n$  trainable parameters (Figure \ref{architectures}(a)), which still fulfills the condition $N_p>\nu$. Also, those results will be compared against choices of $W(\Vec{\theta})$ with only $n$ (Figure \ref{architectures}(b),(d),(e)) and $3\times n$ (Figure \ref{architectures}(c)) trainable parameters per layer, which do not fulfill the condition $N_p>\nu$. \\

The objective of this work is to compare the performance of different choices of $W(\Vec{\theta})$, then the encoding part is the same for all the tests and is described in the following equation,

\begin{equation}
\bigotimes_{m=1}^{M} R_y(x_m) .
\label{encoding}
\end{equation}

This is repeated \textit{vertically} $L$ times in each of the $L$ layers of the circuit, as illustrated in Figure \ref{sp}.\\


The 1D quantum convolution presented in Figure \ref{1Dquanvolution} is utilized in this work. The sliding window size for the experiments in this work is $k=2$  and $s$ varies for each dataset. This effectively allows the circuit to capture the dependencies between adjacent points. The architecture of the circuit depicted in Figure \ref{sp} requires $k \times L$ qubits. This is an advantage of utilizing quantum convolution instead of the quantum version of a multilayer perceptron in which $s\times L$ qubits would be required if the \textit{super parallel} architecture is followed. \\

 The \textit{super parallel} architecture is analyzed with 2, 3 and 4 layers. With 2 layers and $k=2$, 4 qubits are required.  With 3 and 4 layers  6 and 8 qubits are required, respectively. For those architectures, the degree of the output Fourier series is expected to be $D=4,6,8$ respectively. However, depending on the particular choice of $W(\theta)$ some of the coefficients might be set to zero, limiting the expressivity of the model. The observable $R_z$ is measured in all  qubits at the end of the circuit, and the expected value (a real number) of each qubit is passed to the classical part. The output features of the 1D quantum convolution is a matrix with dimensions $(n, o)$, with $o=(s+2\times p-k)/r+1$. Here, $p$ is the padding and $r$ is the stride. In this case, the sample chain of size $s$ is padded at both edges with a zero. The chain is swept with a stride $r=1$. The classical part consists of a ReLU activation function, a max pooling over the dimension $o$ of the output of the quantum convolution, and a linear layer that maps from $n$ elements to finally output the predicted point of the time series. \\

\subsection{Data}
\textcolor{black}{
In this work, we analyze the following datasets: as a toy example, we use the third-order Legendre polynomial with randomly seeded noise; as a small-scale dataset, we consider the USD to EUR exchange rate \cite{ER}; and for more complex and realistic cases, we examine the SP500 \citep{SP500} and Bitcoin \citep{bitcoin} datasets.}

The dataset of the  third Legendre polynomial consist in points generated by the equation  $ P_3(x)=\frac{1}{2}(3x^2-1)$. Random seeded noise was added in each point. The subsequences introduced in  the 1D convolution kernel are of 5 points. 750 points are for training and 250 for testing.\\

The exchange rate data between USD and EUR obtained from \citep{ER} contains daily observations  from January 1, 2020, to July 8, 2021. The model is constructed with 376 simulation data points, given by the sequence $[x(t - 4), x(t - 3), x(t - 2), x(t - 1), x(t); x(t + 1)]$ for $t$ ranging from 5 to 380. The first 300 data points are designated for training, and the remaining points are reserved for testing.\\

\textcolor{black}{
The dataset of Bitcoin \citep{bitcoin} and SP500 \citep{SP500} have data from the open value. The bitcoin dataset has a temporality of 1 minute and the SP500 considers daily observations. In both cases sequences of 5 points to predict the point number 6. In the training stage, 3000 points were were considered and 750 points for testing.}

\section{Results}\label{sec2}
All results are assessed by comparing the root mean squared error (RMSE), mean absolute error (MAE) and mean percentage error (MAPE). These are standard metrics for forecasting models and since they are related to error, a smaller value is indicative of a better result. \\

In this section the following cases are presented. First, the relevance of reuploading is analyzed. For this, we utilize the \textit{parallel} architecture described in \cite{fourier2}. The \textit{parallel} architecture is similar to the \textit{super parallel}, but without reuploading the data vertically: The data is only repeated in a series of layers. Also, a comparison between the \textit{parallel} and \textit{super parallel} architecture is presented. Finally, the performances of ansatz in Figure \ref{architectures} are calculated.  Ansatz (c) is known as \textit{Strongly Entangling Layers}, is a template from \textit{Pennylane} inspired by the work presented in \cite{centric}. Ansatz (d) is the template known as \textit{Basic Entangler}, also from \textit{Pennylane}, with a similar structure as Strongly Entangling, but with a rotation only in one direction. Ansatz (e) is generated with the template \textit{Random Layers} of \textit{Pennylane} and seed=1234, which randomly distributes two-qubit gates and rotations in the circuit. The number of random rotations is derived from the second dimension of weights, and the number of CNOT gates is $\frac{1}{3}$ of the number of rotations.  \\

The Fourier coefficients will be computed for the ansatz in Figure \ref{architectures} to assess the expressivity of the circuit in the context of the FS . Also, the expressivity as described in \cite{expressibility} and the variance of the gradient of the cost function as described in \cite{barren} are computed. To calculate the expressivity, the calculations were generated by sampling 5000 states and using a bin size of 75, following the parameters suggested in \cite{expressibility}.  The circuit is initialized with different sets of random weights and the coefficients are computed from the output of the circuit using a discrete Fourier transform. Those coefficients can be plotted, allowing to easily evaluate the richness of the accessible coefficients of each architecture. The analytical expression of the part of the circuit corresponding to the coefficients is highly non-linear and too complex, then a graphical analysis is more suitable to provide an insight into the differences between each choice of the variational part. \\

\textcolor{black}{The computations were executed with the simulator} \texttt{Pennylane}~\citep{pennylane} \textcolor{black}{on the device \textit{default.qubit}, using the} \texttt{jax-jit} \textcolor{black}{interface. Optimization of weights was performed with the} \texttt{backprop} \textcolor{black}{differentiation method. Just-in-time (} \texttt{jit} \textcolor{black}{) compilation was applied to both the circuit function and the optimization steps. The use of} \texttt{jax}, \texttt{jit} \textcolor{black}{, and vectorization enabled execution in competitive times. All calculations were carried out on a GPU RTX 3080 and a Ryzen 7 5700X processor. All experiments were conducted with a learning rate of} $5 \times 10^{-4}$ \textcolor{black}{using the ADAM optimizer}~\citep{adam}\textcolor{black}{. The weights of the classical model were also initialized with the Xavier method}~\citep{xavier}\textcolor{black}{. Each experiment was repeated 20 times with a different seed, and the average result was reported.}

\subsection{Number of necessary trainable parameters}
\begin{figure}[ht]
	\vskip 0.2in
	\centering
	
	\begin{subfigure}[b]{0.45\textwidth}
		\centering
		\includegraphics[width=0.90\linewidth]{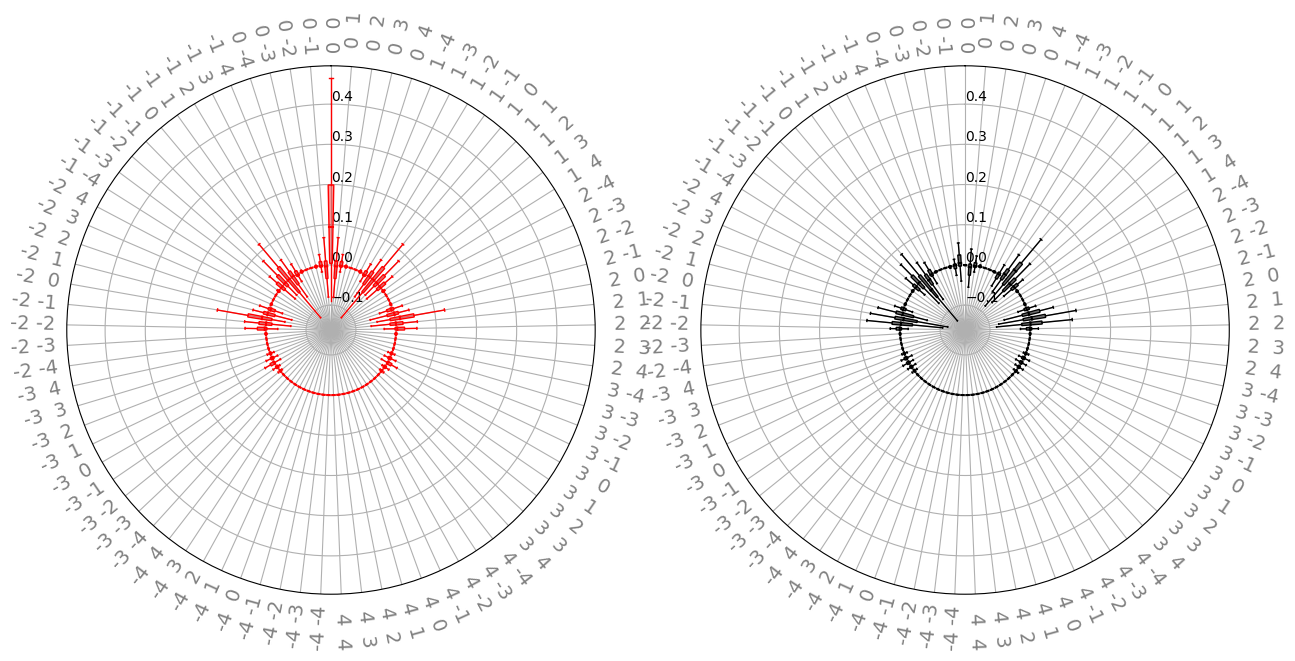}
		\caption{$n\times n=192$ trainable parameters.}
		\label{nxn}
	\end{subfigure}%
	\hfill
	\begin{subfigure}[b]{0.45\textwidth}
		\centering
		\includegraphics[width=0.90\linewidth]{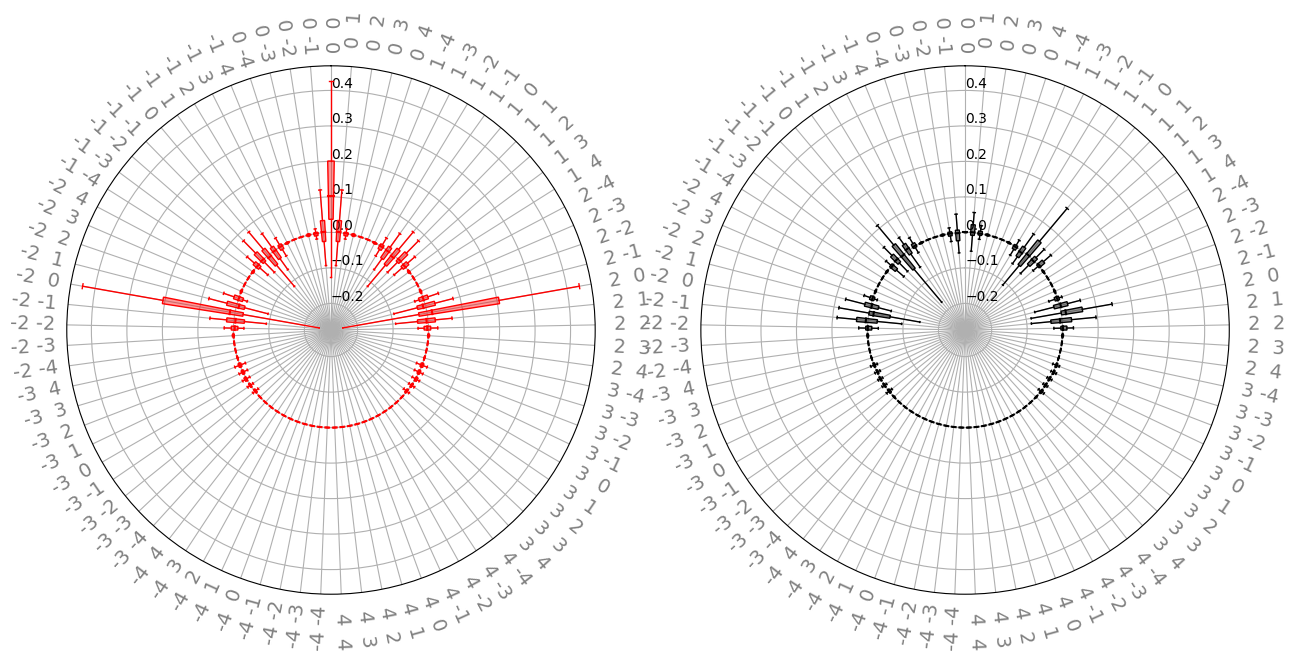}
		\caption{$n\times L=12$ trainable parameters. }
		\label{custom}
	\end{subfigure}
	
	\vskip\baselineskip 
	
	\begin{subfigure}[b]{0.45\textwidth}
		\centering
		\includegraphics[width=0.90\linewidth]{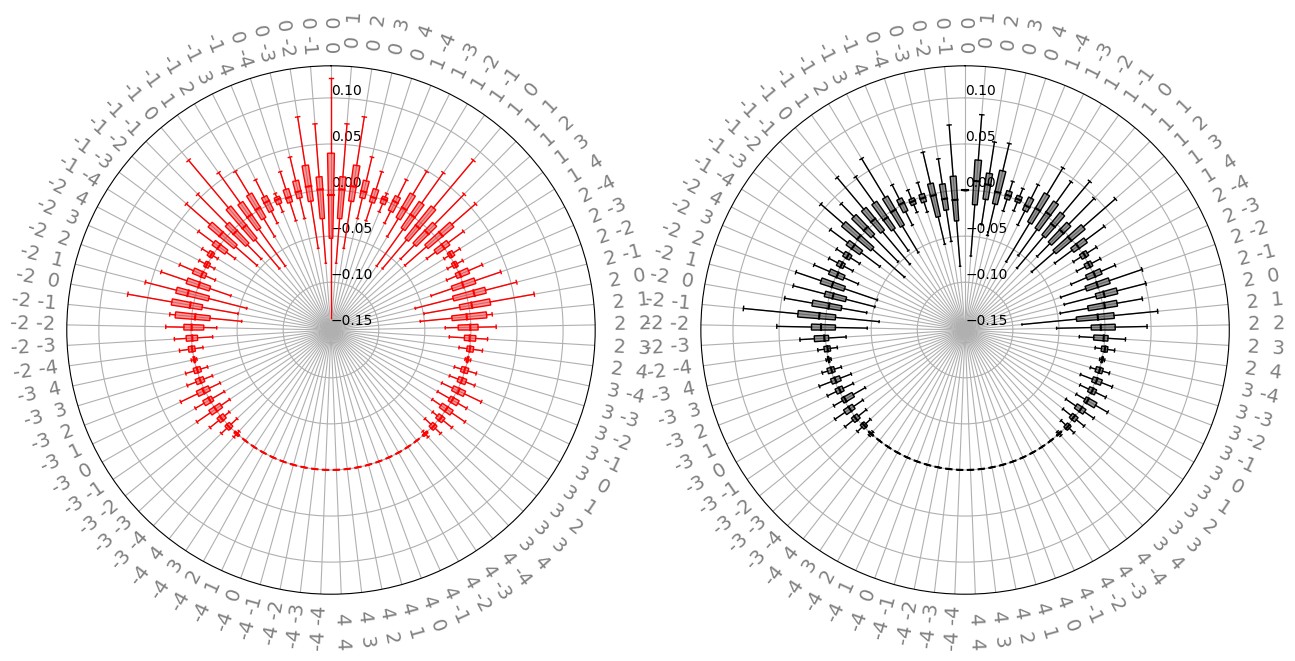}
		\caption{$3\times n \times L=36$ trainable parameters.}
		\label{basic}
	\end{subfigure}%
	\hfill
	\begin{subfigure}[b]{0.45\textwidth}
		\centering
		\includegraphics[width=0.90\linewidth]{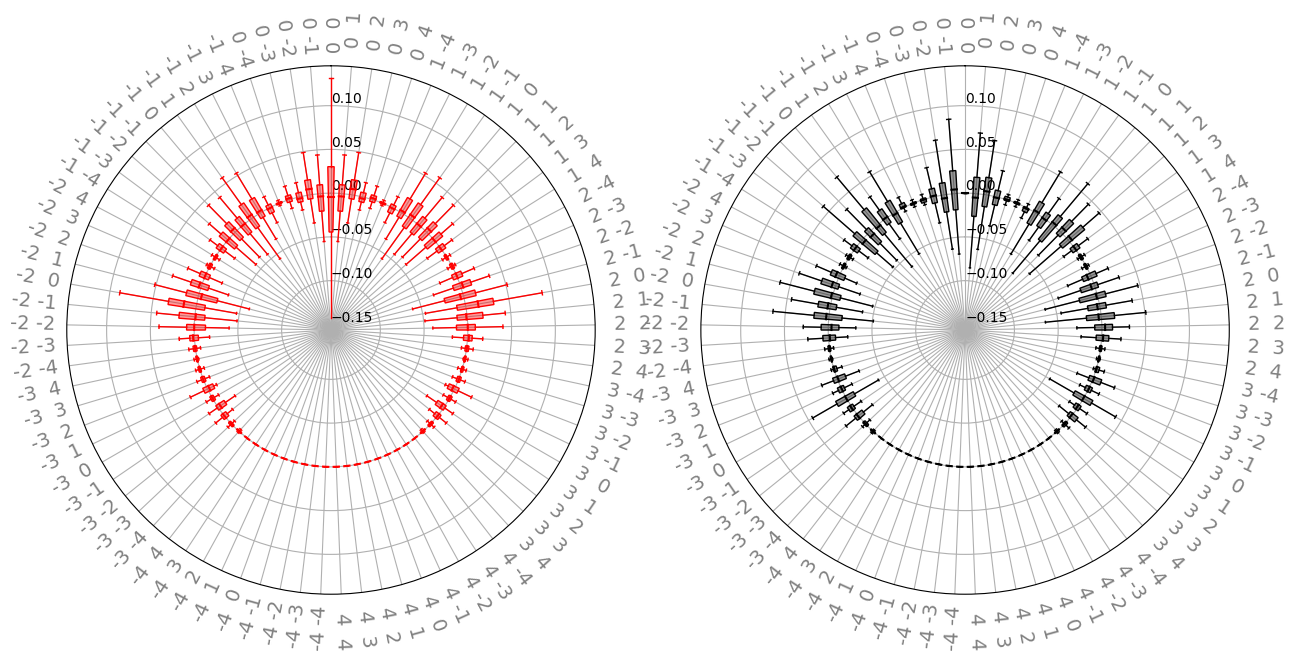}
		\caption{$n\times L=12 $ trainable parameters.}
		\label{strongly}
	\end{subfigure}
	
	\vskip\baselineskip 
	
	\begin{subfigure}[b]{0.45\textwidth}
		\centering
		\includegraphics[width=0.90\linewidth]{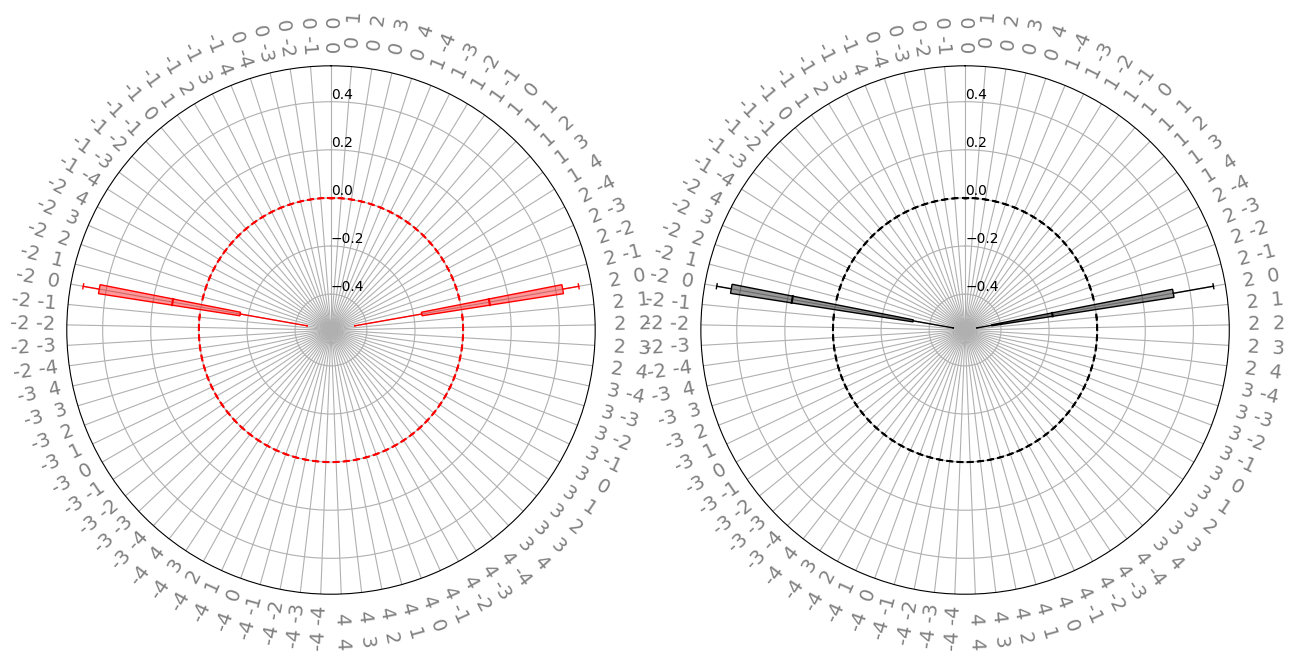}
		\caption{$n \times L=12$ trainable parameters.}
		\label{random}
	\end{subfigure}
	\caption{Accessible coefficients for the architectures of Figure \ref{architectures}. The plots on the left and in red represent the real part. The plots on the right and in black are the imaginary part of the circuit's Fourier coefficients. The labels are the frequencies, with $c_0$ at the top and middle of each circle, and positive frequencies increasing in a clockwise direction and negative frequencies in a counterclockwise direction. The figure was generated by sampling the circuit with 100 different sets of randomly initialized weights. More and bigger bars represent a richer set of values for a coefficient.}
	\label{accessible}
\end{figure}

The sets of accessible coefficients for each of the ansatz in Figure \ref{architectures} are depicted in Figure \ref{accessible}. In this figure, the plots in left and red represent the real part and the right plots in black the imaginary part of the circuit's Fourier coefficients. The labels of the circles are the frequencies. A bigger bar represents a greater distribution of values for a coefficient.   Those results were obtained with 100 different sets of randomly initialized weights. It can be observed that even when the architectures (b), (c), and (d) do not fulfill the condition $N_p>\nu$, they are capable of producing non-zero coefficients.  Moreover, it can be noted that the architectures (c) and (d) are capable of producing a richer set of coefficients than (a), which fulfills the condition. The expected degree for all the architectures is 4, then according to Eq. (\ref{nu}), at least 81 free parameters would be needed to have a general enough circuit. It is noticeable that (b) and (d) with as few as 12 trainable parameters and (c) with 36 trainable parameters are capable of producing more non-zero coefficients than expected. (b) is capable of producing coefficients corresponding to a Fourier series of degree $D=2$, which would need 25 free parameters. (c) and (d) have non-zero coefficients corresponding to a degree $D=3$, for which 49 free parameters would be needed.

In practical terms, achieving high-degree Fourier functions with fewer trainable parameters is highly desirable due to the reduction in training times. For the ansatz (b), (c), and (d) of Figure \ref{architectures}, with 4 qubits and 2 layers, training took on average 0.948 $s$ per epoch. On the other hand, ansatz (a) took on average 16.06 $s$ to complete 1 epoch. Also, the circuits with fewer parameters were capable of achieving comparable metrics, to illustrate this, the results for the Legendre dataset are presented in Table \ref{nxnTable}. The other datasets exhibited similar behavior.\\
\FloatBarrier
\begin{table}[ht]
	\centering
	\begin{tabular}{lrrrrr}
		\hline
		\multicolumn{1}{c}{Ansatz} & \multicolumn{1}{c}{\begin{tabular}[c]{@{}c@{}}Trainable\\ parameters\end{tabular}} & \multicolumn{1}{c}{RMSE} & \multicolumn{1}{c}{MAE} & \multicolumn{1}{c}{MAPE} & \multicolumn{1}{c}{\begin{tabular}[c]{@{}c@{}}Time \\ per epoch(s)\end{tabular}} \\ \hline
		(a)                        & 192                                                                                & 0.1440                   & 0.1109                  & 0.2890                   & 16.06                                                                            \\
		Strongly Entangling (c)    & 36                                                                                 & 0.1330                   & 0.1079                  & 0.2942                   & 1.372                                                                            \\
		Basic Entangler (d)        & 12                                                                                 & 0.1333                   & 0.1081                  & 0.2947                   & 0.948                                                                            \\
		Custom Layers (b)          & 12                                                                                 & \textbf{0.1229}          & \textbf{0.0986}         & \textbf{0.2791}          & 2.301                                                                            \\
		Random Layers (e)          & 12                                                                                 & 0.1654                   & 0.1393                  & 0.4580                   & 1.097                                                                            \\ \hline
	\end{tabular}
	\caption{Comparison of results obtained with the ansatz of Figure \ref{architectures} for the Legendre dataset.}
	\label{nxnTable}
\end{table}
\FloatBarrier

The degree $D$ of the Fourier series for the \textit{parallel} architecture is equal to the number of layers $L$ and for the \textit{super-parallel} architecture is given by the square of the layers $L$. When more qubits and layers are added, ansatz (b), (c), and (d) are capable of producing more non-zero terms than expected by their respective number of trainable parameters. This can be appreciated in Table \ref{moreq}: in Column 6, the actual trainable parameters of each case are presented, and in Column 7 the parameters needed to fulfill the condition $N_P>\nu$, being the values in Column 7 significantly larger than in Column 6.

Having a small number of free parameters producing more non-zero degree coefficients than expected is due to the highly coupled and non-linear form of the coefficients equation; the relationship between the parameters and the Fourier coefficients is highly non-linear, allowing the model to capture complex dependencies with a small number of parameters.

\subsection{Relevance of reuploading}

A comparison between the results of the metrics with and without reuploading is presented. In this case, two circuits are compared. For the non-reuploading case,  an initial trainable layer $W^{(1)}$, the encoding of the information of the two points kernel into 2 qubits, followed by 4 layers of the trainable ansatz. This is compared against the \textit{parallel} architecture with 2 qubits and 4 layers.  \\

The values of the expressivity, calculated as proposed in \cite{expressibility}, for each of the trainable layers in Figure \ref{architectures} are calculated for both cases and depicted in Table \ref{expressibililty_nr}. It can be observed that both cases present comparable values, but an analysis of the metrics in the testing stage for each data set shows that the reuploading strategy is superior. This behavior can be explained if the accessible coefficients for each case are analyzed.\\

\begin{figure*}[ht]
\centering
\includegraphics[width=0.90\linewidth]{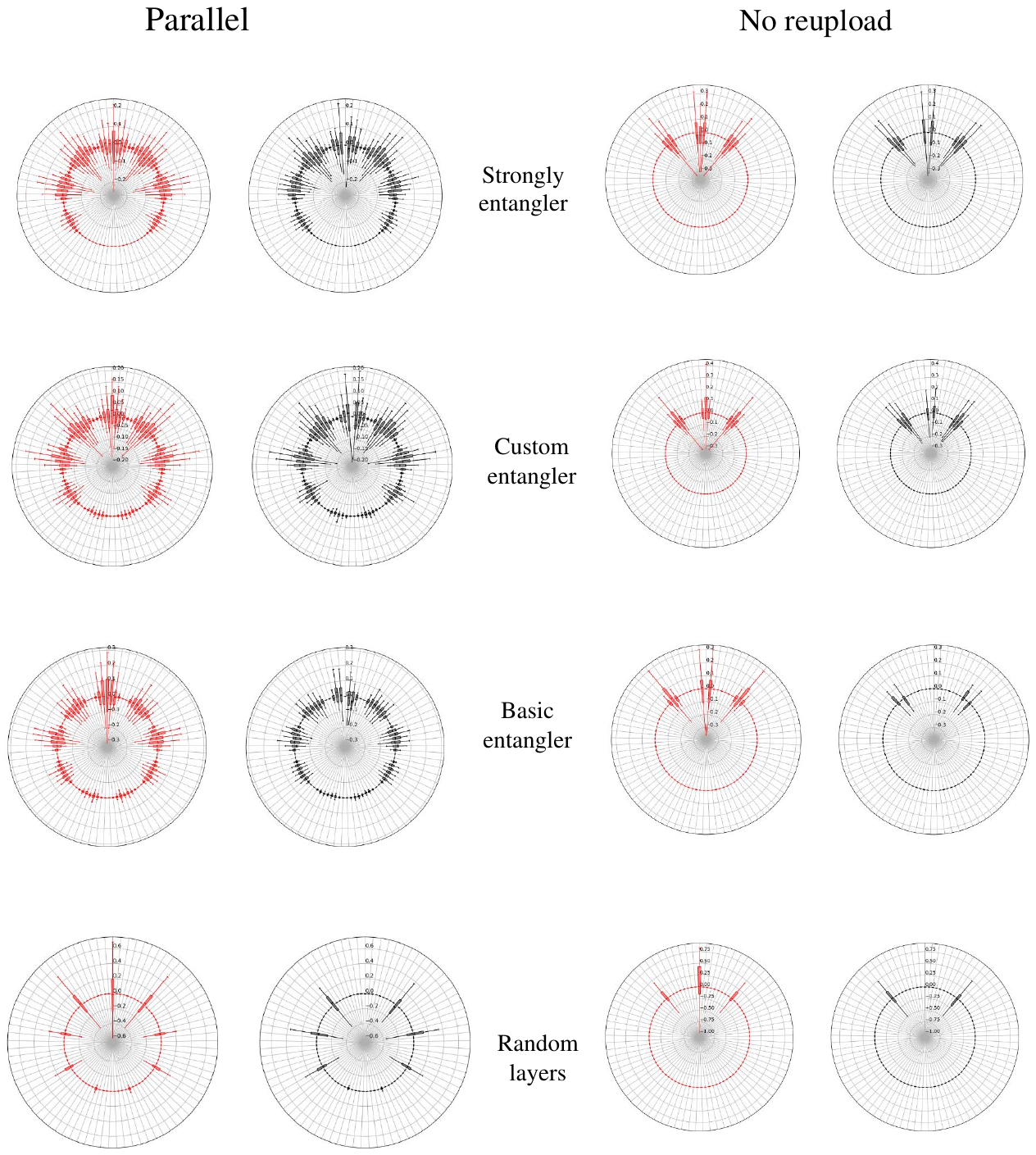}
\caption{Comparison of the accessible coefficients for the parallel and the case without reuploading.  The ansatz are depicted in Figure \ref{architectures}.  As in Figure \ref{accessible}, the plots in right and black are the imaginary part of the circuit's Fourier coefficients. The labels are the frequecies, with $c_0$. The figure was generated by sampling the circuit with  100 different sets of randomly initialized weights. More and bigger bars represents a richer set  of values for a coefficient. It can be clearly observed that more coefficients are accessible to the model when we have reuploading.}
\label{pvsnr}

\end{figure*}

 In Figure \ref{pvsnr} a comparison between the accessible coefficients for the \textit{parallel} and the case without reuploading is presented. It can be observed a clear difference between both architectures, showing that when no reuploading of data is performed, the circuit has significantly less accessible frequencies.  \\

\subsection{\textit{Parallel} vs \textit{super parallel} architectures}
The performance of the \textit{parallel} and \textit{super parallel} architectures presented in \cite{fourier2} is compared. A \textit{super parallel} architecture with 4 qubits and a kernel of size 2 is compared against a \textit{parallel} architecture with 2 qubits, a kernel of 2, and 4 layers. In this way,  the data is re-uploaded the same number of times, and the circuits are expected to output Fourier series of the same degree $d=4$.\\

 In general, as can be observed in Table \ref{pvssp}, the \textit{super parallel} architecture produced better results overall.  In this case, no major difference between the accessible coefficients to both architectures is presented, but as can be noted in Table \ref{pvssp}, the values for the expressivity are better for the \textit{super parallel} case. Even when training times are shorter for the \textit{parallel} case, the difference is not significant enough.  \\

\FloatBarrier
\begin{table}[]
	\begin{tabular}{lrrrrrr}
		\hline
		\multicolumn{1}{c}{Ansatz}           & \multicolumn{1}{c}{Qubits} & \multicolumn{1}{c}{Layers} & \multicolumn{1}{c}{\begin{tabular}[c]{@{}c@{}}Expected\\ degree\end{tabular}} & \multicolumn{1}{c}{\begin{tabular}[c]{@{}c@{}}Obtained\\ degree\end{tabular}} & \multicolumn{1}{c}{\begin{tabular}[c]{@{}c@{}}Trainable \\ parameters\end{tabular}} & \multicolumn{1}{c}{\begin{tabular}[c]{@{}c@{}}Parameters\\ needed for\\ obtained $D$\end{tabular}} \\ \hline
		\multirow{2}{*}{Strongly Entangling} & 6                          & 3                          & 9                                                                             & 8                                                                             & 72                                                                                  & 289                                                                                                \\
		& 8                          & 4                          & 16                                                                            & 11                                                                            & 120                                                                                 & 529                                                                                                \\
		\multirow{2}{*}{Basic Entangler}     & 6                          & 3                          & 9                                                                             & 8                                                                             & 24                                                                                  & 289                                                                                                \\
		& 8                          & 4                          & 16                                                                            & 10                                                                            & 40                                                                                  & 441                                                                                                \\
		\multirow{2}{*}{Custom Layers}       & 6                          & 3                          & 9                                                                             & 5                                                                             & 24                                                                                  & 121                                                                                                \\
		& 8                          & 4                          & 16                                                                            & 8                                                                             & 40                                                                                  & 289                                                                                                \\
		\multirow{2}{*}{Random Layers}       & 6                          & 3                          & 9                                                                             & 2                                                                             & 24                                                                                  & 25                                                                                                 \\
		& 8                          & 4                          & 16                                                                            & 2                                                                             & 40                                                                                  & 25                                                                                                 \\ \hline
	\end{tabular}
	\caption{Degree obtained for 6 and 8 qubits with 3 and 4 layers, respectively. In the last column, the parameters that in theory would be needed to fulfill the condition $N_p>\nu$.}
	\label{moreq}
\end{table}

\begin{table}	
\begin{tabular}{lrrrrr}
	\hline
	\multicolumn{1}{c}{Ansatz} & \multicolumn{2}{c}{Expressivity}                                            & \multicolumn{3}{c}{\begin{tabular}[c]{@{}c@{}}Average $\%$ improvement\\ by using reuploading\end{tabular}} \\ \hline
	\multicolumn{1}{c}{}       & \multicolumn{1}{c}{\textit{Parallel}} & \multicolumn{1}{c}{Non-reuploading} & \multicolumn{1}{c}{RMSE}           & \multicolumn{1}{c}{MAE}           & \multicolumn{1}{c}{MAPE}           \\
	Strongly Entangling        & 0.0071                                & 0.0082                              & 13.77                              & 15.16                             & 28.39                              \\
	Basic Entangler            & 0.0210                                & 0.0375                              & 15.57                              & 17.25                             & 37.11                              \\
	Custom Layers              & 0.0090                                & 0.0100                              & 13.93                              & 16.50                             & 39.67                              \\
	Random Layers              & 0.8729                                & 0.8784                              & 21.99                              & 21.73                             & 7.39                               \\ \hline
\end{tabular}
	
	\caption{Expressivity for the \textit{parallel} architecture vs non-reuploading architecture and average taken over the three datasets of improvement (reduction of RMSE, MAE and MAPE) by using reuploading }
	\label{expressibililty_nr}
\end{table}

\begin{table}[]
\centering
\begin{tabular}{lrrrrrr}
	\hline
	\multicolumn{1}{c}{Ansatz} & \multicolumn{2}{c}{Expressivity}                                                    & \multicolumn{3}{c}{\begin{tabular}[c]{@{}c@{}}Average $\%$ improvement\\ by using \textit\{super parallel\}\end{tabular}} & \multicolumn{1}{c}{}                                                                      \\ \hline
	\multicolumn{1}{c}{}       & \multicolumn{1}{c}{\textit{Parallel}} & \multicolumn{1}{c}{\textit{Super Parallel}} & \multicolumn{1}{c}{RMSE}                     & \multicolumn{1}{c}{MAE}                     & \multicolumn{1}{c}{MAPE}                    & \multicolumn{1}{c}{\begin{tabular}[c]{@{}c@{}}Training time\\ \% difference\end{tabular}} \\
	Strongly Entangling        & 0.0071                                & 0.0033                                      & 10.27                                        & 9.89                                        & 10.53                                       & 17.20                                                                                     \\
	Basic Entangler            & 0.0210                                & 0.0053                                      & 7.77                                         & 10.54                                       & 3.63                                        & 18.21                                                                                     \\
	Custom Layers              & 0.0090                                & 0.0054                                      & 19.86                                        & 17.61                                       & 17.57                                       & 14.98                                                                                     \\
	Random Layers              & 0.8729                                & 2.764                                       & 36.93                                        & 36.59                                       & 19.50                                       & 12.11                                                                                     \\ \hline
\end{tabular}
\caption{Values for the expressivity of the \textit{parallel }and super \textit{parallel} architectures for each different ansatz, along with the corresponding percentage improvements in standard metrics when employing the \textit{super parallel} architecture compared to the \textit{parallel} architecture.}
\label{pvssp}
\end{table}
\FloatBarrier


\subsection{Comparison between ansatz}
\begin{figure*}[]
	\vskip 0.2in
	\centering
	
	\begin{subfigure}[b]{0.70\textwidth}
		\centering
		\includegraphics[width=0.70\linewidth]{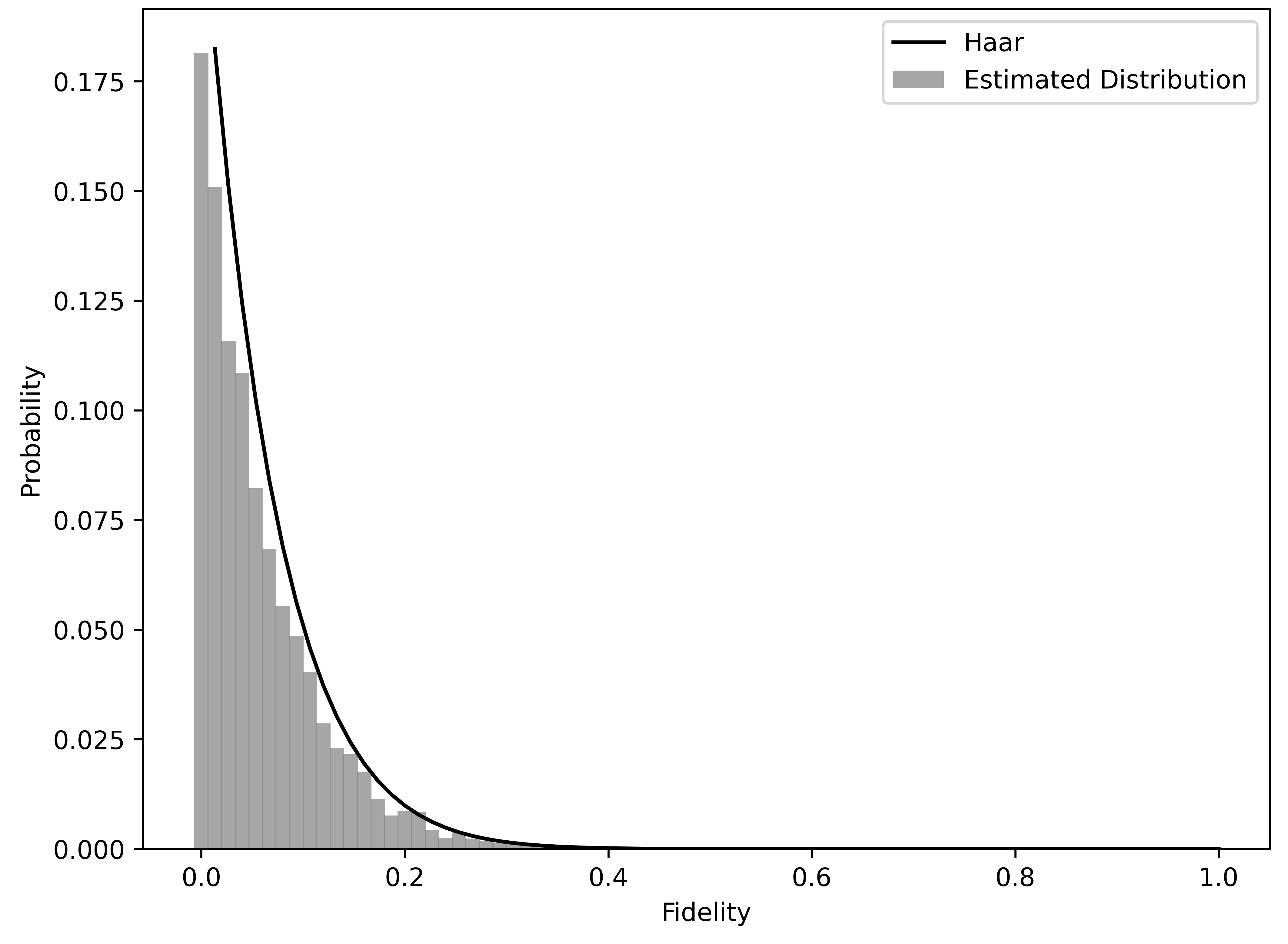}
		\caption{Expressivity (KL divergence) =0.0033.}
		\label{hist_rand}
	\end{subfigure}%
	
	\vskip\baselineskip 
	
	\begin{subfigure}[b]{0.70\textwidth}
		\centering
		\includegraphics[width=0.70\linewidth]{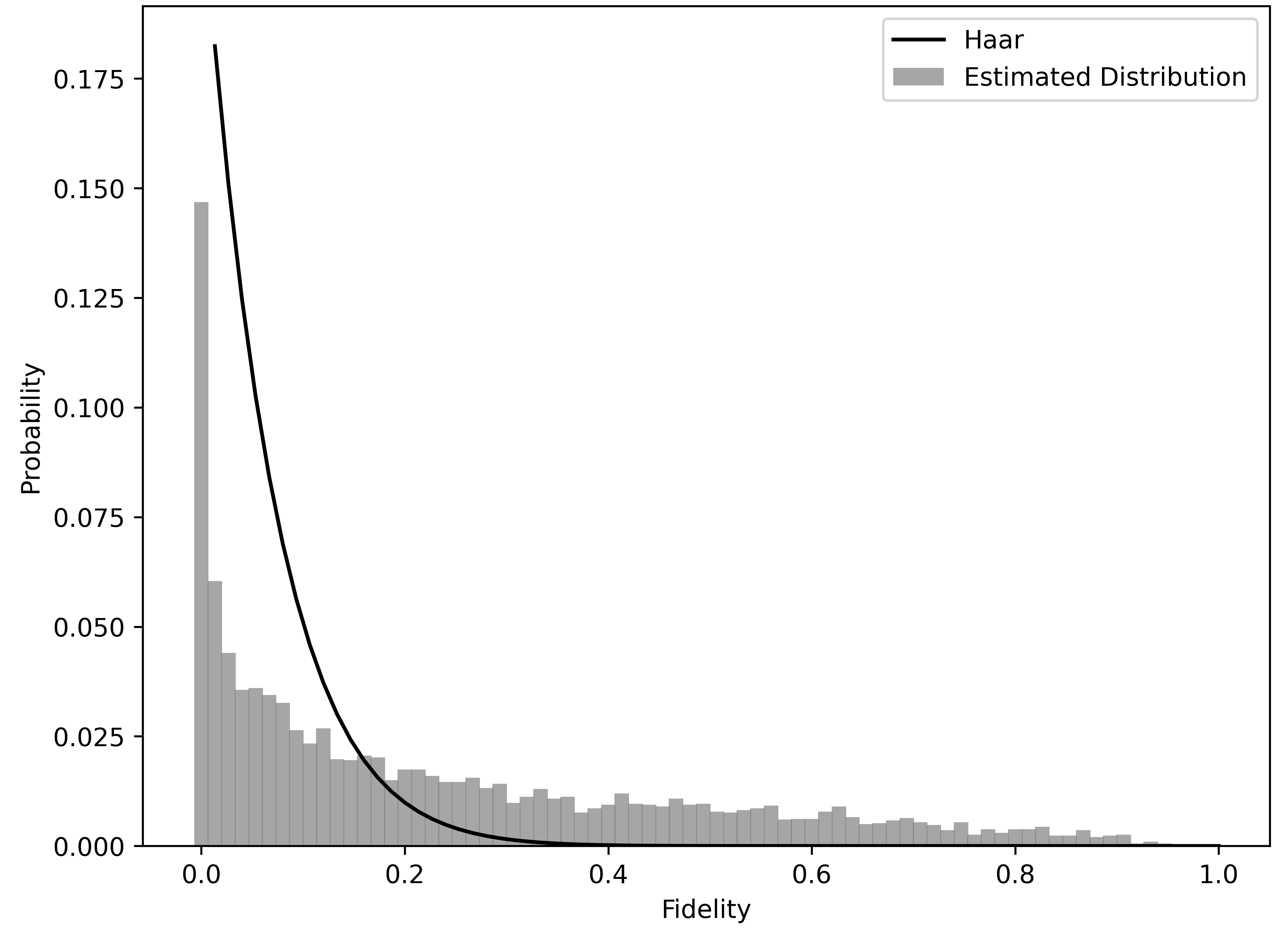}
		\caption{Expressivity (KL divergence) =2.764}
		\label{hist_strong}
	\end{subfigure}
	\caption{Histograms of estimated
		fidelities, overlaid with fidelities of the Haar-distributed ensemble. Values in subcaptions are calculated as explained in Subsection 2.5 and in \cite{expressibility} }
	\label{histogram}
\end{figure*}

The ansatz (b)-(e) in Figure \ref{architectures} are tested for 4, 6, and 8 qubits with 2, 3, and 4 layers respectively. In Table \ref{expr} the values of the degree of the obtained Fourier function, the expressivity, and the variance of the derivative of the cost function are presented. It can be observed a relation between the accessible coefficients and the expressivity, both parameters are complementary. The ansatz (e) has only 2 accessible coefficients for the 3 tested cases and also presents a high value of the Kulbak-Leibler divergence, significantly different from (b)-(d). Visually, no major differences can be observed when comparing the histograms of the calculated fidelities overlaid with the Haar fidelities for (b)-(d) for the different number of qubits, but it is illustrative to compare the histograms for the less expressible ansatz (Random Layers (e)) and the most expressible ansatz (Strongly Layers (c)). In Figure \ref{histogram} this comparison is presented for the case with 4 qubits. A clear difference between the calculated and theoretical fidelities is observed for the case with low expressivity in the Random Layers ansatz (Figure \ref{hist_rand}) and a closer proximity between both is appreciated for the case with high expressivity with the Strongly Layers ansatz (Figure \ref{hist_strong}). As it can be noted from Figure \ref{architectures}, the difference between this ansatz and the three others is the number of entangling CNOT gates; while the other cases have all their qubits entangled, in (e) only $\frac{1}{3}$ of the qubits are entangled. The results for (b)-(d) are comparable, but some differences are evident. The Basic Entangler and the Strongly Entangling have a similar structure: each qubit is rotated first and after that, all qubits are entangled. This similarity results in the closeness of the three analyzed values;  those are also slightly different from Custom Layers  (which has a different structure), especially in the manner that the values evolve when adding more qubits. This impact of the structure of an ansatz can also be noted in Figure \ref{accessible}: (a) and (d) have the same accessible coefficients despite the difference of trainable parameters (192 and 12, respectively) and the same happens for (b) and (c). The results for (b) are superior to (c). This is easily explainable since (b) performs a rotation in the 3 directions, being able to cover the Hilbert space in a more complete form. Also, (b) has three times more trainable parameters than (c). The expressivity and the accessible coefficients for (b) grow when adding more qubits, but this also implies a lower value for the variance of the derivative of the cost function, which might result in a flat cost landscape.  Ansatz (d) presents lower values for expressivity and has less accessible coefficients than (b) and (c), but on the other hand the value of the variance evolves slowly, implying better trainability. \\

\begin{table*}[]
\centering
\begin{tabular}{lrrrrr}
	\hline
	\multicolumn{1}{c}{Ansatz}           & \multicolumn{1}{c}{Qubits} & \multicolumn{1}{c}{Layers} & \multicolumn{1}{c}{\begin{tabular}[c]{@{}c@{}}Obtained\\ degree\end{tabular}} & \multicolumn{1}{c}{Expressivity } & \multicolumn{1}{c}{\begin{tabular}[c]{@{}c@{}}Variance of \\ derivative\end{tabular}} \\ \hline
	\multirow{3}{*}{Strongly Entangling} & 4                          & 2                          & 3                                                                             & 0.0033                             & 0.021                                                                                 \\
	& 6                          & 3                          & 8                                                                             & 0.0013                             & 0.0063                                                                                \\
	& 8                          & 4                          & 11                                                                            & 0.00011                            & 0.0016                                                                                \\ \hline
	\multirow{3}{*}{Basic Entangler}     & 4                          & 2                          & 3                                                                             & 0.0053                             & 0.0434                                                                                \\
	& 6                          & 3                          & 8                                                                             & 0.0008                             & 0.0132                                                                                \\
	& 8                          & 4                          & 10                                                                            & 0.0004                             & 0.0032                                                                                \\ \hline
	\multirow{3}{*}{Custom Layers}       & 4                          & 2                          & 2                                                                             & 0.0054                             & 0.0454                                                                                \\
	& 6                          & 3                          & 5                                                                             & 0.0075                             & 0.0223                                                                                \\
	& 8                          & 4                          & 8                                                                             & 0.0032                             & 0.0112                                                                                \\ \hline
	\multirow{3}{*}{Random Layers}       & 4                          & 2                          & 2                                                                             & 2.764                              & 0.1270                                                                                \\
	& 6                          & 3                          & 2                                                                             & 5.604                              & 0.0860                                                                                \\
	& 8                          & 4                          & 2                                                                             & 1.737                              & 0.0605                                                                                \\ \hline
\end{tabular}
\caption{Degree, expressivity, and variance of the derivative of the cost function for the ansatz (b)-(e) in Figure \ref{architectures}. The cases with 4, 6, and 8 qubits are analyzed. }
\label{expr}
\end{table*}

\begin{figure*}[t!]
\centering
\includegraphics[width=0.8\linewidth]{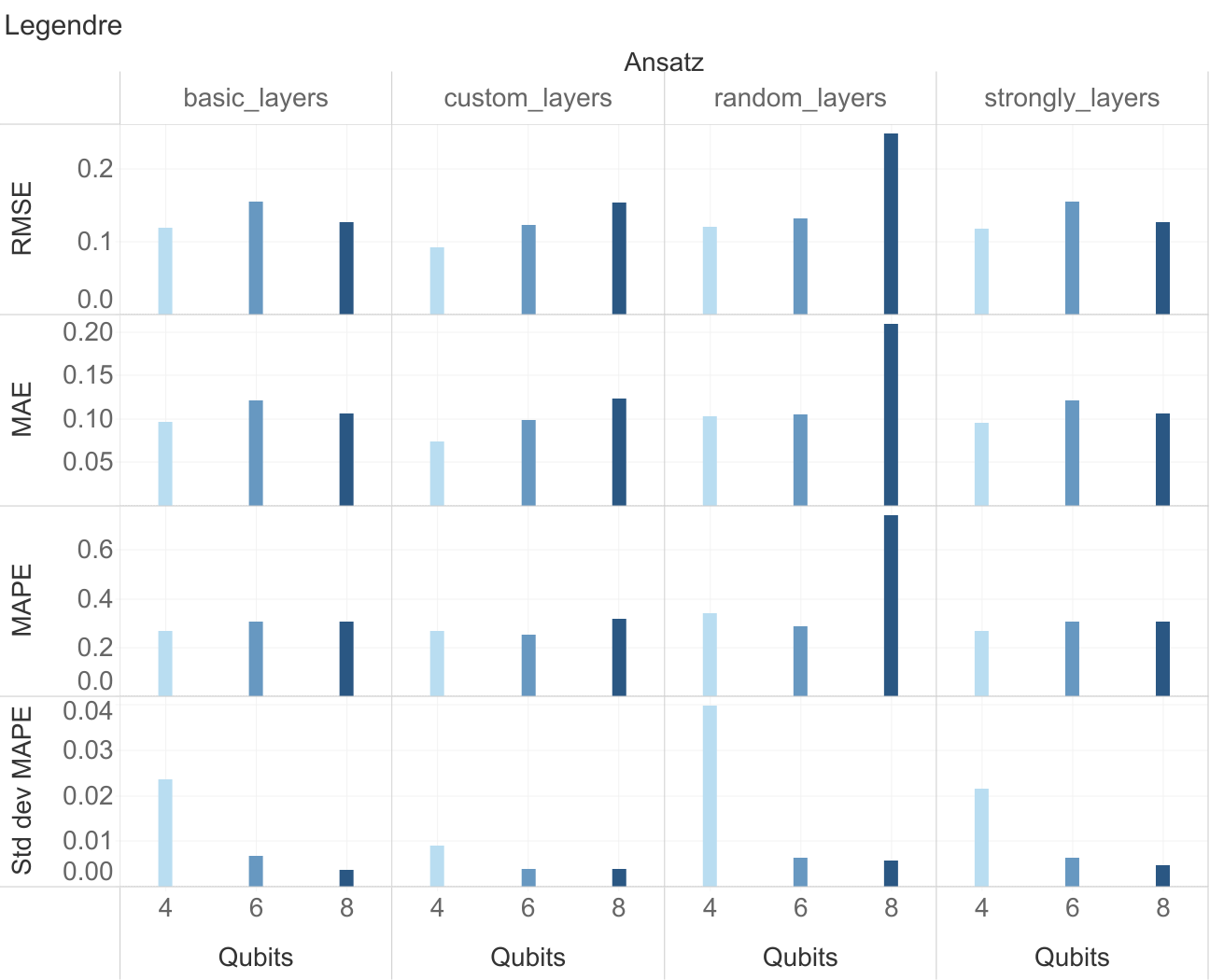}
\caption{Comparison between ansatz for the Legendre polynomial dataset 4, (light blue) 6, and 8 (dark blue) qubits. The values for RMSE, MAE, and MAPE are presented, then a smaller bar represents a better result.  Results are similar but the 8 qubits configuration achieves a lower standar deviation.}
\label{legendre8q}
\end{figure*}

\begin{figure*}[t!]
\centering
\includegraphics[width=0.8\linewidth]{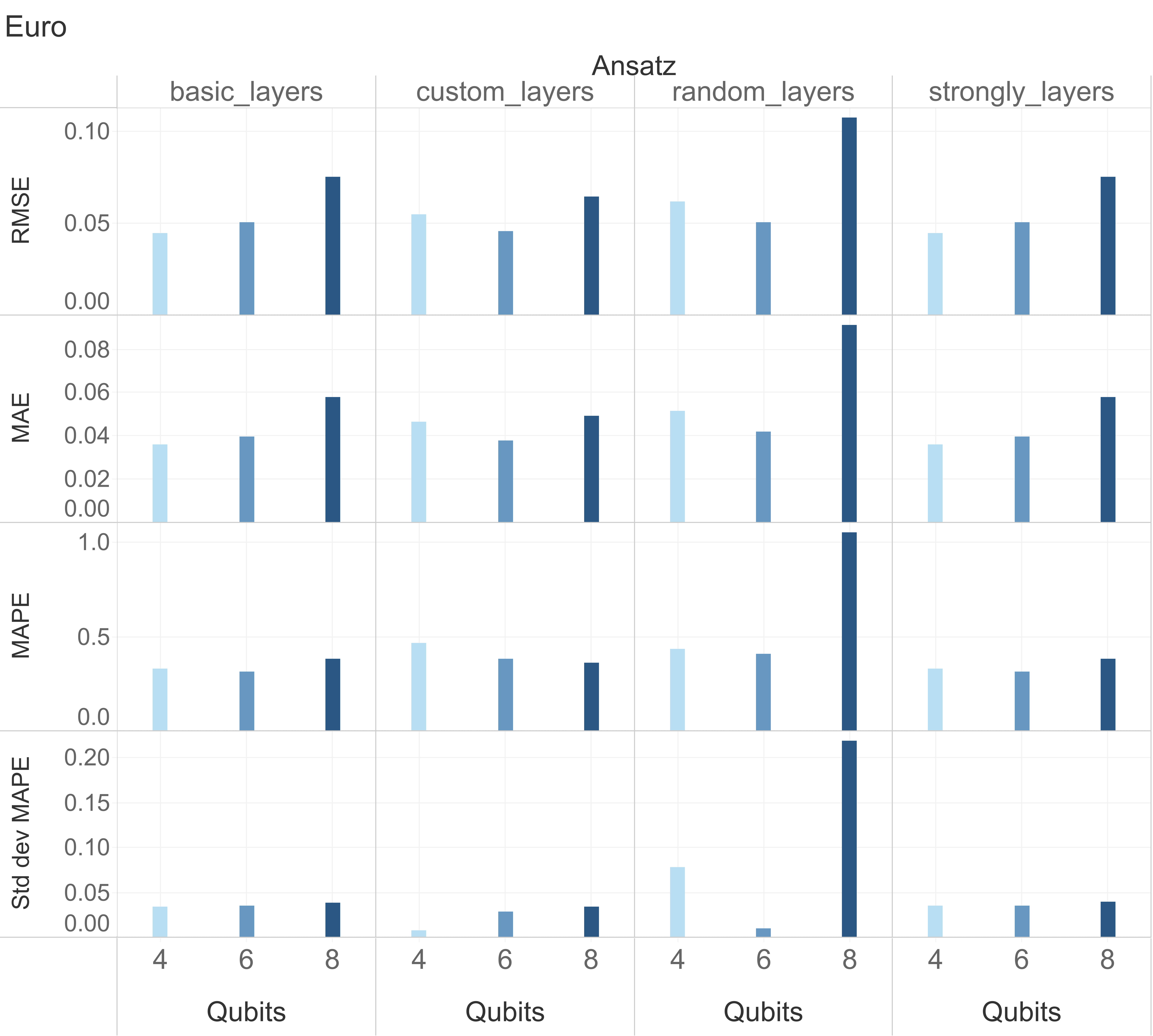}
\caption{Comparison between 4, (light blue) 6, and 8 (dark blue) qubits for the Euro dataset. The values for RMSE, MAE, MAPE and the standard deviation of MAPE are presented.}
\label{euro_ansatz}
\end{figure*}

\begin{figure*}[t!]
	\centering
	\includegraphics[width=0.8\linewidth]{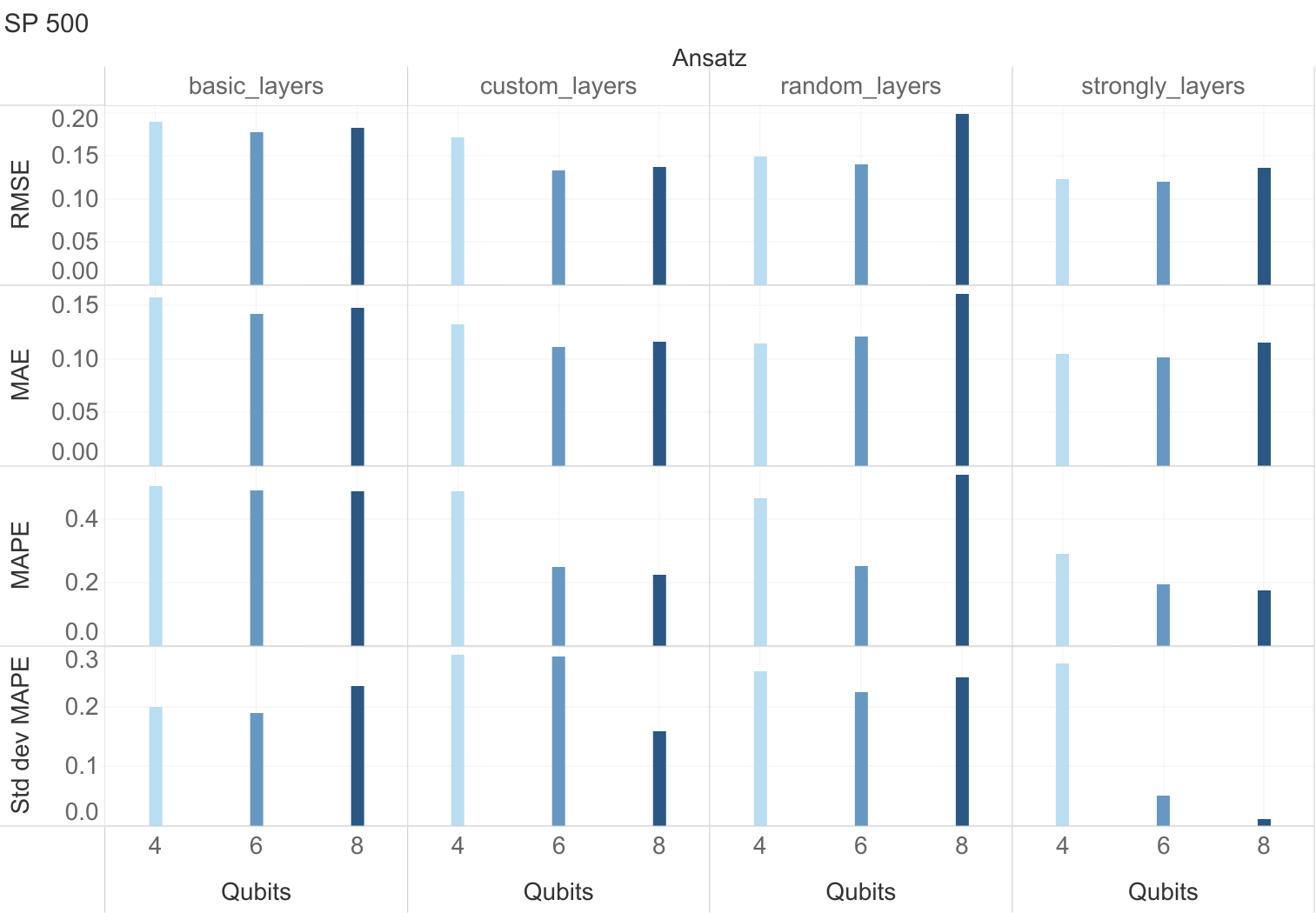}
	\caption{Comparison between ansatz for the SP500 dataset 4, (light blue) 6, and 8 (dark blue) qubits. The values for RMSE, MAE, MAPE, and the standard deviation of MAPE are presented.  The 8 qubits configuration achieves a lower standar deviation.}
	\label{sp500}
\end{figure*}

\begin{figure*}[t!]
	\centering
	\includegraphics[width=0.8\linewidth]{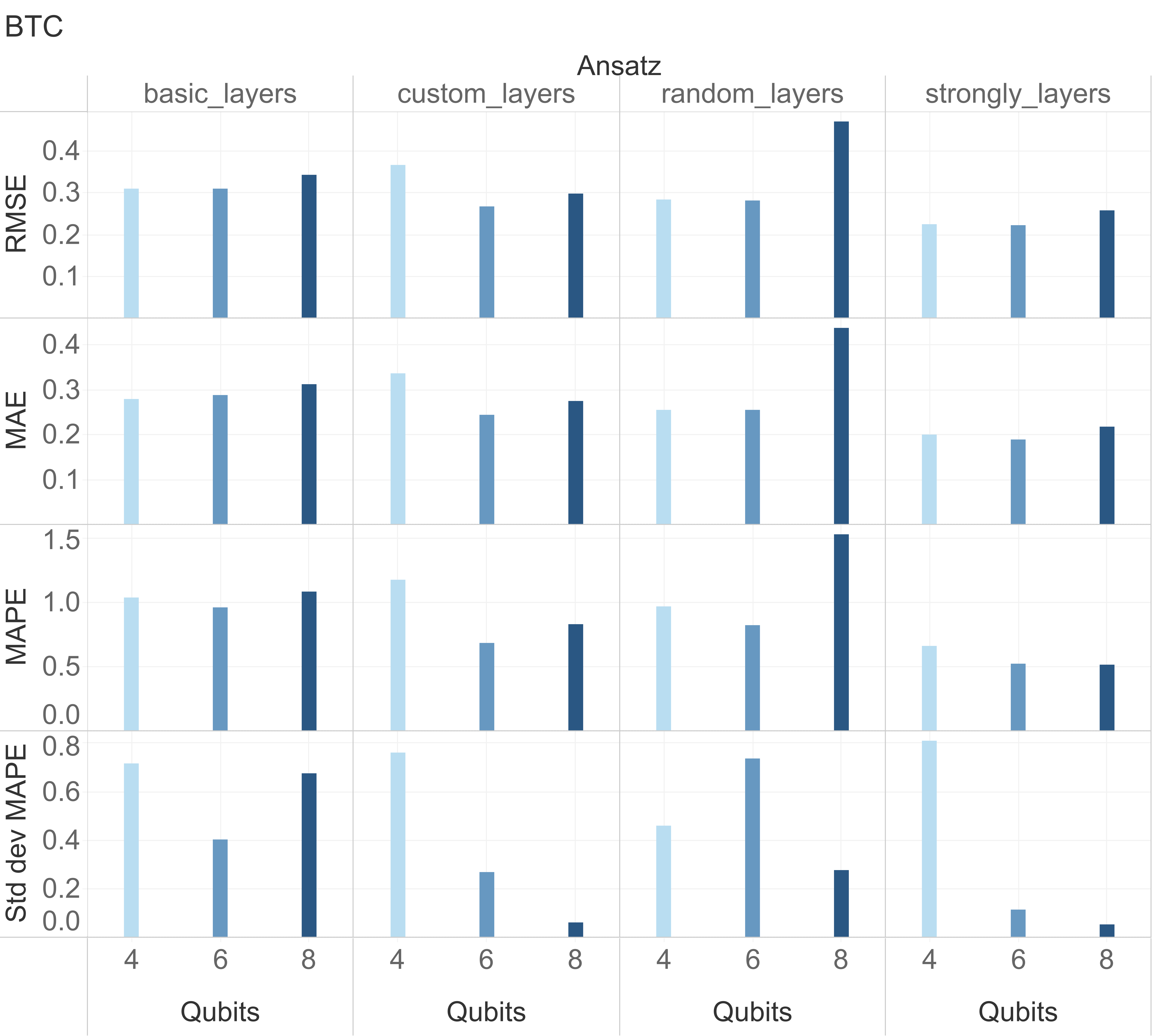}
	\caption{Comparison between ansatz for the BTC dataset 4, (light blue) 6, and 8 (dark blue) qubits. The values for RMSE, MAE, MAPE, and the standard deviation of MAPE are presented.  The 8 qubits configuration achieves a lower standar deviation.}
	\label{btc}
\end{figure*}

Testing the impact of the obtained degree values, expressiveness, and variance with real data is important. For this purpose, the datasets described in Subsection 3.1 are utilized with the hybrid model described in Section 3. Those results are depicted in the bar plots of Figures \ref{legendre8q} to \ref{btc}. In those plots the obtained metrics for 4 (light blue), 6, and 8 (dark blue) qubits and for each of the ansatz of Figure \ref{architectures}. Since the metrics RMSE, MAE and MAPE are related to the error, a smaller bar represents a better result. \\

\textcolor{black}{
The Legendre dataset is an example of a simple dataset. This case is presented in Figure \ref{legendre8q}. Similar results are observed for RMSE, MAE and MAPE by Basic Layers, Custom Layers and Strongly Layers in the 4, 6 and 8 qubits configurations. The Random Layers ansatz configuration with 4 and 6 qubits achieved comparable but consistently resulted in higher errors than the other three ansatz. This result is expected given its low expressibility. A considerably higher error was obtained with the Random Layers and 8 qubits. In general, it can be observed in the standard deviation pannel that the 8 qubits configuration consistently achieved a more stable result.}

\textcolor{black}{
 The Euro dataset, trained with only 300 samples,  is an example of a small dataset. This result is presented in Figure \ref{euro_ansatz}. In the Euro dataset, the values of the metrics seem to be more sensitive to the results of Table \ref{expr}. In this case, differently than in the example of a simple dataset, by observing the standar deviation pannel the 8 qubits configuration is not the most stable case. It can be observed that metrics are better both in stability and overall result with 4 qubits. This result can be explained because even when the values of the expressivity increase with the number of qubits, this comes with a decrease in the divergence of the gradient of the cost function. \\}

\textcolor{black}{
The impact of the expressive power of an ansatz is better appreciated in a more complex dataset. In Figures \ref{sp500} and \ref{btc} the Strongly Layers ansatz consistently achieved better metrics, an expected result given its high expressivity. It can also be observed that results are considerably more stable for the 8 qubit configuration with Strongly Layers.
}

\textcolor{black}{
	Across the three datasets, the Random Layers ansatz resulted in slightly less favorable metrics, as anticipated due to its limited expressivity and absence of non-zero Fourier coefficients. For the toy dataset, the results were comparable across ansatz, but a more stable result is achieved with 8 qubits. The small dataset proved to be more challenging to train using the most expressive ansatz. In the more complex datasets (SP500 and BTC) the expressivity of the Strongly Entangler with 8 qubits is key to obtain a good prediction.
}

\begin{figure*}[t!]
	\centering
	\includegraphics[width=0.99\linewidth]{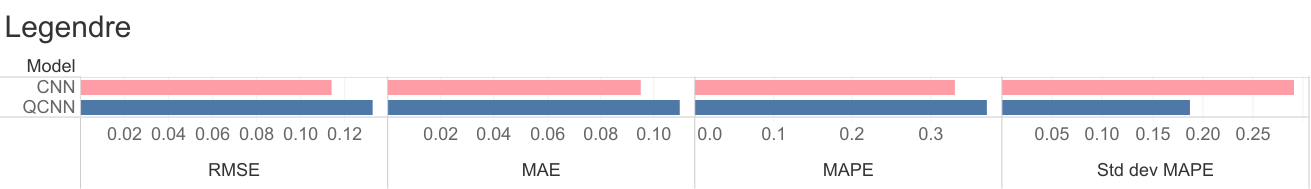}
	\caption{Comparison between the classical counterpart of our model (pink bars) and the 1D QCNN (blue bars). In each case, the results are averaged over implementations using 4, 6, and 8 output channels. For the quantum model, the average includes all results obtained with the four previously analyzed ansatz. The metrics obtained are comparable for both models. }
	\label{qvsc}
\end{figure*}

\textcolor{black}{
To complete our analysis, Figure \ref{qvsc} presents a comparison between our quantum model and its classical counterpart using the Legendre dataset. For both models, we report the average performance using 4, 6, and 8 output channels, with each configuration evaluated across 20 different random seeds. In the quantum case, the reported results are also averaged across all previously considered ansatz. The metrics obtained are generally comparable: the classical model achieves slightly better performance in terms of RMSE, MAE, and MAPE, while the QCNN shows a smaller standard deviation. 
}

\textcolor{black}{
It is important to note that, since our study focused on a simplified architecture with only a single convolutional layer, the reported errors are not optimal. In a previous work by our research group \citep{mayra}, we performed a similar comparison between a 1D QCNN and its classical version, aiming to obtain the best possible metrics in both cases. That earlier study used more complex architectures, including three convolutional layers and two fully connected layers. In contrast, the present work prioritizes simplicity to better analyze the behavior of the quantum layer and understand the impact of quantum layer design. Therefore, the results shown in Figures \ref{euro_ansatz} to \ref{qvsc} are not the lowest achievable errors; better performance would likely be obtained by adding more layers, as done in our previous study.
}

\section{Conclusion}\label{sec13}

In this study, we take the theoretical insights from \cite{fourier, fourier2,expressibility,barren} to evaluate the performance of four different ansatz, considering practicality in training with real datasets. Numerous ansatz and qubit combinations exist and exploring all cases is beyond the scope of this work. Also, the results may vary with different datasets. Nevertheless, some key conclusions can be drawn from our findings.\\

Contrary to the condition emphasized in \cite{fourier2} that $N_p>\nu$ is necessary, our results suggest that a limited number of trainable parameters can be capable to produce Fourier functions of higher degrees. This is an indicative of the remarkable expressive power of quantum circuits. This observation is particularly relevant in terms of training times and also offers insight into why previous works achieved good results with few parameters. Further work needs to be done to provide more information about why few trainable parameters are capable of producing higher degree Fourier series than expected.\\

By analyzing the problem of quantum 1D convolution in the context of Fourier transforms, it was possible to design the architecture in a form in which data is reuploaded (\textit{parallel} and \textit{super parallel} architectures), which lead to significantly improved results. This improvement can be infered from the theory presented in \cite{fourier} and \cite{fourier2}, where they also pointed time series as a natural application to their work.  Employing a \textit{super parallel} structure proves more effective than reuploading the data an equivalent number of times in a \textit{parallel} structure. The significance of data reuploading might not have been evident solely from an expressivity analysis:  it only becomes apparent when considering accessible coefficients, and the superiority of the \textit{super parallel} approach cannot be inferred from the coefficients, but is noted when analyzing the expressivity. Then, it can be said that both parameters are complementary, this emphasizes the importance of a comprehensive analysis to obtain an optimal ansatz.\\

Regarding specific ansatz performances, Strongly Entangler, Custom Entangler, and Basic Entangler consistently gave favorable results, with a tendency to obtain \textcolor{black}{a more stable result}. The expressivity of Strongly Entangling also comes with the price of having a flat optimization landscape when increasing the number of qubits. When analyzing the testing metrics, it can be seen that this  affected the Euro dataset, which had fewer training points. \textcolor{black}{In the cases of complex datasets, the expressivity of the ansatz was more important}. These findings contribute valuable insights for selecting effective ansatz configurations in quantum machine learning applications. 

\backmatter

\bmhead{Supplementary information}

The data, code, and results are available at \url{https://github.com/SandraJuarez/1D-Quantum-Convolutional-Neural-Network}.

\bmhead{Aknowledgements}
The authors acknowledge the support of Consejo Nacional de Humanidades, Ciencias y Tecnologias. Also, the thechnical support and comments of JI Hernandez-Martinez
\section*{Declarations}

\begin{itemize}

\item \textbf{Conflict of interest/Competing interests:} The authors declare no conflicts of interest. The authors have no relevant financial or non-financial interests to disclose. All authors certify that they have no affiliations with or involvement in any organization or entity with any financial interest or non-financial interest in the subject matter or materials discussed in this manuscript.The authors have no financial or proprietary interests in any material discussed in this article.
\item Ethics approval and consent to participate: Ethics approval Not applicable as this study did not involve any human.
or animal subjects.
\item \textbf{Author contribution:}  S.L.J.O. designed and conducted the experiments, analyzed the obtained data and wrote the manuscript. M.A.R.R. designed the 1D quantum convolution and the corresponding code and proposed the datasets. A.M.V, E.R.T. and J.M.L.R. provided supervision, guidence and resources throughout the development of the project. Aditionally, M.A.R.R., A.M.V and E.R.T. provided critical insights that shaped the direction of the research. All authors reviewed the manuscript.

\item \textbf{Funding} Open Access funding enabled by CINVESTAV. This study is  supported by the Consejo Nacional de Humanidades, Ciencias y Tecnologias, CONAHCYT

\item \textbf{Consent for publication}
\end{itemize}

\noindent


\bibliography{references}

\begin{thebibliography}{10}
\providecommand{\doi}[1]{\url{https://doi.org/#1}}
\bibcommenthead

\bibitem[\protect\citeauthoryear{Cerezo et~al.}{2021}]{variational}
Cerezo M, Arrasmith A, Babbush R, Benjamin SC, Endo S, Fujii K, et~al.
\newblock Variational quantum algorithms.
\newblock Nature Reviews Physics. 2021;3(9):625--644.

\bibitem[\protect\citeauthoryear{De~Luca}{2021}]{surveyQML}
De~Luca G.
\newblock Survey of NISQ Era Hybrid Quantum-Classical Machine Learning
  Research.
\newblock Journal of Artificial Intelligence and Technology. 2021
  12;\doi{10.37965/jait.2021.12002}.

\bibitem[\protect\citeauthoryear{Li and Deng}{2021}]{review}
Li W, Deng DL.
\newblock Recent advances for quantum classifiers.
\newblock Science China Physics, Mechanics and Astronomy. 2021 dec;65(2).
\newblock \doi{10.1007/s11433-021-1793-6}.

\bibitem[\protect\citeauthoryear{Henderson et~al.}{2020}]{henderson}
Henderson M, Shakya S, Pradhan S, Cook T.
\newblock Quanvolutional neural networks: powering image recognition with
  quantum circuits.
\newblock Quantum Machine Intelligence. 2020;2(1):2.

\bibitem[\protect\citeauthoryear{Schuld et~al.}{2020}]{PhysRevA.101.032308}
Schuld M, Bocharov A, Svore KM, Wiebe N.
\newblock Circuit-centric quantum classifiers.
\newblock Phys Rev A. 2020 Mar;101:032308.
\newblock \doi{10.1103/PhysRevA.101.032308}.

\bibitem[\protect\citeauthoryear{Park et~al.}{2023}]{park2022variational}
Park G, Huh J, Park DK.
\newblock Variational quantum one-class classifier.
\newblock Machine Learning: Science and Technology. 2023 jan;4(1):015006.
\newblock \doi{10.1088/2632-2153/acafd5}.

\bibitem[\protect\citeauthoryear{Mari et~al.}{2020}]{abejas}
Mari A, Bromley TR, Izaac J, Schuld M, Killoran N.
\newblock Transfer learning in hybrid classical-quantum neural networks.
\newblock {Quantum}. 2020 Oct;4:340.
\newblock \doi{10.22331/q-2020-10-09-340}.

\bibitem[\protect\citeauthoryear{Sameer and Gupta}{2022}]{sameer2022novel}
Sameer M, Gupta B.
\newblock A Novel Hybrid Classical-Quantum Network to Detect Epileptic
  Seizures.
\newblock medRxiv. 2022;p. 2022--05.

\bibitem[\protect\citeauthoryear{Houssein et~al.}{2022}]{10.1093/jcde/qwac003}
Houssein EH, Abohashima Z, Elhoseny M, Mohamed WM.
\newblock {Hybrid quantum-classical convolutional neural network model for
  COVID-19 prediction using chest X-ray images}.
\newblock Journal of Computational Design and Engineering. 2022
  02;9(2):343--363.
\newblock \doi{10.1093/jcde/qwac003}.
\newblock
  {\href{https://arxiv.org/abs/https://academic.oup.com/jcde/article-pdf/9/2/343/42616958/qwac003.pdf}{{https://academic.oup.com/jcde/article-pdf/9/2/343/42616958/qwac003.pdf}}}.

\bibitem[\protect\citeauthoryear{Shahwar et~al.}{2022}]{alzhimer}
Shahwar T, Zafar J, Almogren A, Zafar H, Rehman A, Shafiq M, et~al.
\newblock Automated Detection of Alzheimer’s via Hybrid Classical Quantum
  Neural Networks.
\newblock Electronics. 2022 02;11:721.
\newblock \doi{10.3390/electronics11050721}.

\bibitem[\protect\citeauthoryear{Yang and Sun}{2022}]{defects}
Yang YF, Sun M.
\newblock Semiconductor Defect Detection by Hybrid Classical-Quantum Deep
  Learning.
\newblock In: 2022 {IEEE}/{CVF} Conference on Computer Vision and Pattern
  Recognition ({CVPR}). {IEEE}; 2022. Available from:
  \url{https://doi.org/10.11092Fcvpr52688.2022.00236}.

\bibitem[\protect\citeauthoryear{Alejandra et~al.}{2022}]{time_series}
Alejandra RRM, Andres MV, Mauricio LRJ.
\newblock Time Series Forecasting with Quantum Machine Learning Arquitectures.
\newblock In: Obdulia Pichardo~Lagunas BMS Juan Martínez-Miranda, editor.
  Advances in Computational Intelligence; 2022. p. "66--82".

\bibitem[\protect\citeauthoryear{Rivera-Ruiz et~al.}{2024}]{mayra}
Rivera-Ruiz MA, Ju{\'a}rez-Osorio SL, Mendez-Vazquez A, L{\'o}pez-Romero JM,
  Rodriguez-Tello E.
\newblock 1D Quantum Convolutional Neural Network for Time Series Forecasting
  and Classification.
\newblock In: Calvo H, Mart{\'i}nez-Villase{\~{n}}or L, Ponce H, editors.
  Advances in Computational Intelligence. Cham: Springer Nature Switzerland;
  2024. p. 17--35.

\bibitem[\protect\citeauthoryear{Hur et~al.}{2022}]{qconvolution}
Hur T, Kim L, Park DK.
\newblock Quantum convolutional neural network for classical data
  classification.
\newblock Quantum Machine Intelligence. 2022 feb;4(1).
\newblock \doi{10.1007/s42484-021-00061-x}.

\bibitem[\protect\citeauthoryear{Hong et~al.}{2021}]{hong2021quantum}
Hong Z, Wang J, Qu X, Zhu X, Liu J, Xiao J.
\newblock Quantum convolutional neural network on protein distance prediction.
\newblock In: 2021 International Joint Conference on Neural Networks (IJCNN).
  IEEE; 2021. p. 1--8.

\bibitem[\protect\citeauthoryear{Schuld et~al.}{2021}]{fourier}
Schuld M, Sweke R, Meyer JJ.
\newblock Effect of data encoding on the expressive power of variational
  quantum-machine-learning models.
\newblock Physical Review A. 2021 mar;103(3).
\newblock \doi{10.1103/physreva.103.032430}.

\bibitem[\protect\citeauthoryear{Casas and Cervera-Lierta}{2023}]{fourier2}
Casas B, Cervera-Lierta A.
\newblock Multidimensional Fourier series with quantum circuits.
\newblock Phys Rev A. 2023 Jun;107:062612.
\newblock \doi{10.1103/PhysRevA.107.062612}.

\bibitem[\protect\citeauthoryear{Sim et~al.}{2019}]{expressibility}
Sim S, Johnson PD, Aspuru-Guzik A.
\newblock Expressibility and Entangling Capability of Parameterized Quantum
  Circuits for Hybrid Quantum-Classical Algorithms.
\newblock Advanced Quantum Technologies. 2019;2(12):1900070.
\newblock \doi{https://doi.org/10.1002/qute.201900070}.
\newblock
  {\href{https://arxiv.org/abs/https://onlinelibrary.wiley.com/doi/pdf/10.1002/qute.201900070}{{https://onlinelibrary.wiley.com/doi/pdf/10.1002/qute.201900070}}}.

\bibitem[\protect\citeauthoryear{Holmes et~al.}{2022}]{barren}
Holmes Z, Sharma K, Cerezo M, Coles PJ.
\newblock Connecting Ansatz Expressibility to Gradient Magnitudes and Barren
  Plateaus.
\newblock PRX Quantum. 2022 Jan;3:010313.
\newblock \doi{10.1103/PRXQuantum.3.010313}.

\bibitem[\protect\citeauthoryear{Feynman}{2018}]{feynman2018simulating}
Feynman RP.
\newblock Simulating physics with computers.
\newblock In: Feynman and computation. CRC Press; 2018. p. 133--153.

\bibitem[\protect\citeauthoryear{Preskill}{2023}]{40years}
Preskill J.
\newblock Quantum computing 40 years later.
\newblock Nature Reviews Physics. 2023 jan;4(1).
\newblock \doi{https://doi.org/10.1038/s42254-021-00410-6}.

\bibitem[\protect\citeauthoryear{Nielsen and Chuang}{2000}]{nielsen}
Nielsen MA, Chuang IL.
\newblock Quantum Computation and Quantum Information.
\newblock Cambridge University Press; 2000.

\bibitem[\protect\citeauthoryear{Stoudenmire and
  Schwab}{2016}]{stoudenmire2016supervised}
Stoudenmire E, Schwab DJ.
\newblock Supervised learning with tensor networks.
\newblock In: Advances in Neural Information Processing Systems. vol.~29; 2016.
  p. 4799--4807.

\bibitem[\protect\citeauthoryear{Huggins et~al.}{2019}]{huggins2019towards}
Huggins W, Patil P, Mitchell B, Whaley KB, Stoudenmire EM.
\newblock Towards quantum machine learning with tensor networks.
\newblock Quantum Science and Technology. 2019;4(2):024001.

\bibitem[\protect\citeauthoryear{Chen et~al.}{2022}]{chen2022residual}
Chen Y, Pan Y, Dong D.
\newblock Residual tensor train: A quantum-inspired approach for learning
  multiple multilinear correlations.
\newblock IEEE Transactions on Neural Networks and Learning Systems.
  2022;33(9):4573--4587.

\bibitem[\protect\citeauthoryear{Schuld and
  Petruccione}{2018}]{libro_supervised}
Schuld M, Petruccione F.
\newblock Supervised Learning with Quantum Computers.
\newblock Springer; 2018.

\bibitem[\protect\citeauthoryear{Schuld et~al.}{2020}]{centric}
Schuld M, Bocharov A, Svore KM, Wiebe N.
\newblock Circuit-centric quantum classifiers.
\newblock Physical Review A. 2020 Mar;101(3).
\newblock \doi{10.1103/physreva.101.032308}.

\bibitem[\protect\citeauthoryear{Bergholm et~al.}{2022}]{penny}
Bergholm V, Izaac J, Schuld M, Gogolin C.: PennyLane: Automatic differentiation
  of hybrid quantum-classical computations.

\bibitem[\protect\citeauthoryear{Schuld et~al.}{2019}]{shift}
Schuld M, Bergholm V, Gogolin C, Izaac J, Killoran N.
\newblock Evaluating analytic gradients on quantum hardware.
\newblock Physical Review A. 2019 mar;99(3).
\newblock \doi{10.1103/physreva.99.032331}.

\bibitem[\protect\citeauthoryear{Di~Matteo}{2021}]{measeure}
Di~Matteo O.: Understanding the Haar Measure.
\newblock Xanadu.
\newblock Available from:
  \url{https://pennylane.ai/qml/demos/tutorial_haar_measure}.

\bibitem[\protect\citeauthoryear{Werner}{2023}]{ER}
Werner.: Pacific Exchange Rate Service.
\newblock Accessed on 2023-01-20.
\newblock \url{http://fx.sauder.ubc.ca/data.html}.

\bibitem[\protect\citeauthoryear{Wmcginn}{2021}]{SP500}
Wmcginn.: SP500 csv.
\newblock Accessed: 2025-03-28.
\newblock \url{https://www.kaggle.com/datasets/wmcginn/sp500-csv}.

\bibitem[\protect\citeauthoryear{Zieliński}{2017}]{bitcoin}
Zieliński M.: Bitcoin Historical Data.
\newblock Accessed: 2025-03-28.
\newblock Kaggle.

\bibitem[\protect\citeauthoryear{Bergholm et~al.}{2018}]{pennylane}
Bergholm V, Izaac J, Schuld M, Gogolin C, Blank C, McKiernan K, et~al.
\newblock PennyLane: Automatic differentiation of hybrid quantum-classical
  computations.
\newblock arXiv preprint arXiv:181104968. 2018;.

\bibitem[\protect\citeauthoryear{Kingma and Ba}{2014}]{adam}
Kingma DP, Ba J.
\newblock Adam: A Method for Stochastic Optimization.
\newblock arXiv preprint arXiv:14126980. 2014;.

\bibitem[\protect\citeauthoryear{Glorot and Bengio}{2010}]{xavier}
Glorot X, Bengio Y.
\newblock Understanding the difficulty of training deep feedforward neural
  networks.
\newblock In: Proceedings of the Thirteenth International Conference on
  Artificial Intelligence and Statistics (AISTATS). JMLR Workshop and
  Conference Proceedings; 2010. p. 249--256.

\end{thebibliography}




\end{document}